\documentclass{aa}

\usepackage{graphicx}
\usepackage[skip=0pt]{subcaption}
\usepackage[colorlinks=true,linkcolor=blue,citecolor=blue, urlcolor=blue]{hyperref}
\usepackage{txfonts}

\begin{document} 

    \title{Influence of the large-scale structure velocity field on halo mass assembly within cosmic filaments}

    \titlerunning{Influence of the large-scale structure velocity field on halo mass assembly}

   \author{D. Díaz-Axtle
          \inst{\ref{Leiden}}
          \and
          P. Awad\inst{\ref{Leiden}}
          \and
          D. Rana\inst{\ref{Leiden}}
          \and
          H. Hoekstra\inst{\ref{Leiden}}
          }

   \institute{
    Leiden Observatory, Leiden University, PO Box 9513, NL-2300 RA Leiden, the Netherlands
    \label{Leiden}
    \newline
    \email{dannaeaxtle@hotmail.com}
    }

    \date{Received \today; accepted }

  \abstract{
  The cosmic web governs the growth of dark matter halos through anisotropic matter flows and the resulting large-scale density structure, yet the role of local velocity-field dynamics in halo mass assembly remains poorly understood. The cosmic velocity field encodes non-linear information about matter transport and may influence halo growth independently of local density, particularly within filamentary environments. We investigate the connection between halo mass and the local velocity field, focusing on divergence and vorticity, in order to assess the role of cosmic dynamics in halo mass assembly independently of local density. We analyze dark matter halos in the IllustrisTNG TNG50-1-Dark simulation at $z=0$. Cosmic filaments are extracted using the 1-DREAM framework and divided into three density regimes. We then characterize the velocity field within and around filaments, and study statistical correlations between halo mass, velocity divergence, vorticity, and radial distance from filament spines. The velocity field exhibits a strong correlation with the cosmic web, with converging flows tracing filamentary structures and vorticity arising primarily in overdense, non-linear regions. Within filaments, both divergence and vorticity show enhanced kinematic activity at intermediate distances ($\sim0.5-1.2$~Mpc) from the spine, particularly in intermediate- and high-overdensity filaments, with amplitudes increasing with overdensity. We confirm a clear mass segregation, with more massive halos preferentially located near filament centers. Most notably, halos with masses $M \geq 10^{12}~M_{\odot}$ are found almost exclusively in regions of low absolute divergence and low vorticity, while highly dynamical environments are dominated by low-mass halos. These trends persist across all filament density regimes. We further find that halos residing in  kinematically calm regions assembled later and maintain higher recent accretion rates than halos in dynamically active environments, linking present-day velocity-field properties to halo growth histories. These results demonstrate that the local kinematic state of the cosmic flow encodes information about halo mass and assembly beyond that contained in density alone, providing a complementary dynamical description of halo growth within the cosmic web.
    }
   \maketitle

\section{Introduction}

    On megaparsec scales, matter in the Universe is distributed in a complex hierarchical network known as the cosmic web \citep{bond1996}. The cosmic volume is dominated by vast voids, while anisotropic gravitational collapse leads matter to sequentially form sheets, filaments, and clusters, as described by the Zel’dovich approximation of structure formation \citep{zeldovich1970, peebles1980}. In general, clusters and voids correspond to overdense and underdense cosmic environments, respectively, while filaments and sheets occupy a wide range of densities \citep{cautun2014}. Together, these asymmetric structures form the backbone of the cosmic web and have been shown to play a significant role in influencing the properties of both galaxies and the dark matter (DM) halos which contain them \citep{ChenEtal2017, KraljicEtal2020, CastignaniEtal2022, MalavasiEtal2022, BulichiEtal2023, RajEtal2024, StorckEtal2025, YuEtal2025, ZarattiniEtal2025,TobarEtal2026}.  
    
    Cosmic filaments constitute an intermediate-density environment characterized by collapse along two spatial axes and expansion along the third \citep{zeldovich1970}. The nature of this collapse renders elongated structures with quasi-linear distributions of DM halos that extend from a few to hundreds of megaparsecs \citep{aragon2010}. While the hierarchical assembly of halos is a global feature of $\Lambda$CDM, the specific geometry of filaments provides a unique laboratory for testing the details of halo formation \citep{StorckEtal2025, HadzhiyskaEtal2025, ShimEtal2026}. 

    Despite the strong correlation between environment and halo properties, our understanding of the precise mechanisms driving halo mass build-up remains incomplete. Classical models of structure formation describe halo formation in terms of the collapse of overdense regions in the linear density field, effectively treating local density as the primary predictor of halo growth \citep[e.g.][]{press1974, Sommerville1999, ColeEtal2000}. While this approach successfully captures many statistical properties of the halo population, it relies on the idea that matter accretion is encoded solely in the density field. However, halo formation is inherently a dynamical process, and recent studies suggest that density alone does not fully capture its complexity. Notably, \cite{etezad2025} have shown that the inclusion of information regarding the cosmic velocity field at early redshift improves the prediction of the halo mass function in the present, therefore hinting that the velocity field plays a role in regulating mass assembly, particularly within the complex, non-linear regimes of the large-scale structure. Velocity-space components, such as divergence and vorticity, encode the infalling of matter and the directional nature of accretion, providing a more granular description of the flow of matter in the Universe and supplementing density-based studies.

    Furthermore, the cosmic velocity field is a direct dynamical tracer of structure formation by encoding the response of matter to gravitational collapse. While the velocity field is well-constrained by linear perturbation theory in low-density and in mildly non-linear regimes \citep{peebles1980, bernardeau2002}, filaments represent a critical regime where this mapping breaks down. In these evolved environments, significant non-linear effects, shell-crossing, and vorticity generation \citep{laigle2015} create complex flow patterns that cannot be captured by linear theory alone. Accurately characterizing these non-linear cosmic flows therefore relies on high-resolution N-body simulations to explore mass transport and assembly within them. The link between halo formation and the cosmic velocity field, in addition to being under-explored in the literature (mostly limited to studies of the origin of halo angular momenta), is now more relevant than ever with the advent of wide-field spectroscopic surveys, such as \textit{DESI} \citep{desi2016}, through its dedicated Peculiar Velocity (PV) Survey \citep{desipv2023}, and the \textit{4MOST Hemisphere Survey of the Nearby Universe} (4HS; \citealt{4most2023}). These surveys will provide complementary coverage of the northern and southern skies, enabling increasingly complete reconstructions of the large-scale velocity field and measurements of structure growth in the nearby Universe. In this context, incorporating velocity-field information into studies of halo formation offers a promising avenue to better constrain the physical processes driving mass assembly. 

    In this work, we investigate the possible connections between halo mass and the large-scale velocity field to further shed the light on the role of the latter in halo mass assembly. We focus on regions characterized by strong inflows/outflows (encoded in the divergence of the velocity field) and by swirling motion (or vorticity). To test if the relations between halo mass and the dynamical field are not actually driven by local density, we focus our study on cosmic filaments divided between three density bins. With that aim, we utilize the Illustris-TNG Dark simulations \citep{IllustrisTNG} at redshift $z=0$, and apply the 1-DREAM (1-Dimensional Recovery, Extraction, and Analysis of Manifolds) toolbox \citep{canducci20221dream} for the extraction of cosmic filaments within the simulation box and their further study. 1-DREAM is an interpretable machine learning framework specifically designed to identify and model 1D structures from noisy, high-dimensional data. The toolbox was originally validated as a proof-of-concept for characterizing filamentary topology \citep{canducci20221dream} and subsequently utilized by \cite{petra20231dream} to measure local properties of cosmic filaments (e.g. DM particle density within cross-sections of filaments) and to differentiate the various cosmic web environments, with results comparable to the state-of-the-art tools compiled in \cite{libeskind2018}. This work represents its first application to the IllustrisTNG simulated data.
    
    This paper is organized as follows. Section \ref{sec:data} introduces the numerical simulations and data set employed to define the cosmic volume for our analysis; this is followed by Section \ref{sec:methodology}, where we introduce 1-DREAM and our procedure for extracting cosmic filaments and the sample of halos within them.   
    We present our results in Section \ref{sec:results}, where we analyze how the properties of the cosmic velocity field within filaments, specifically divergence and vorticity, correlate with halo mass and radial distance from the spine. In Section \ref{sec:discussion}, we discuss how these relationships shift across the three defined overdensity categories as well as the implications of our findings on halo mass assembly. Section \ref{sec:conclusions} concludes our work and suggests future extensions.
    
   \begin{figure}
   \centering
    \includegraphics[width=\hsize]{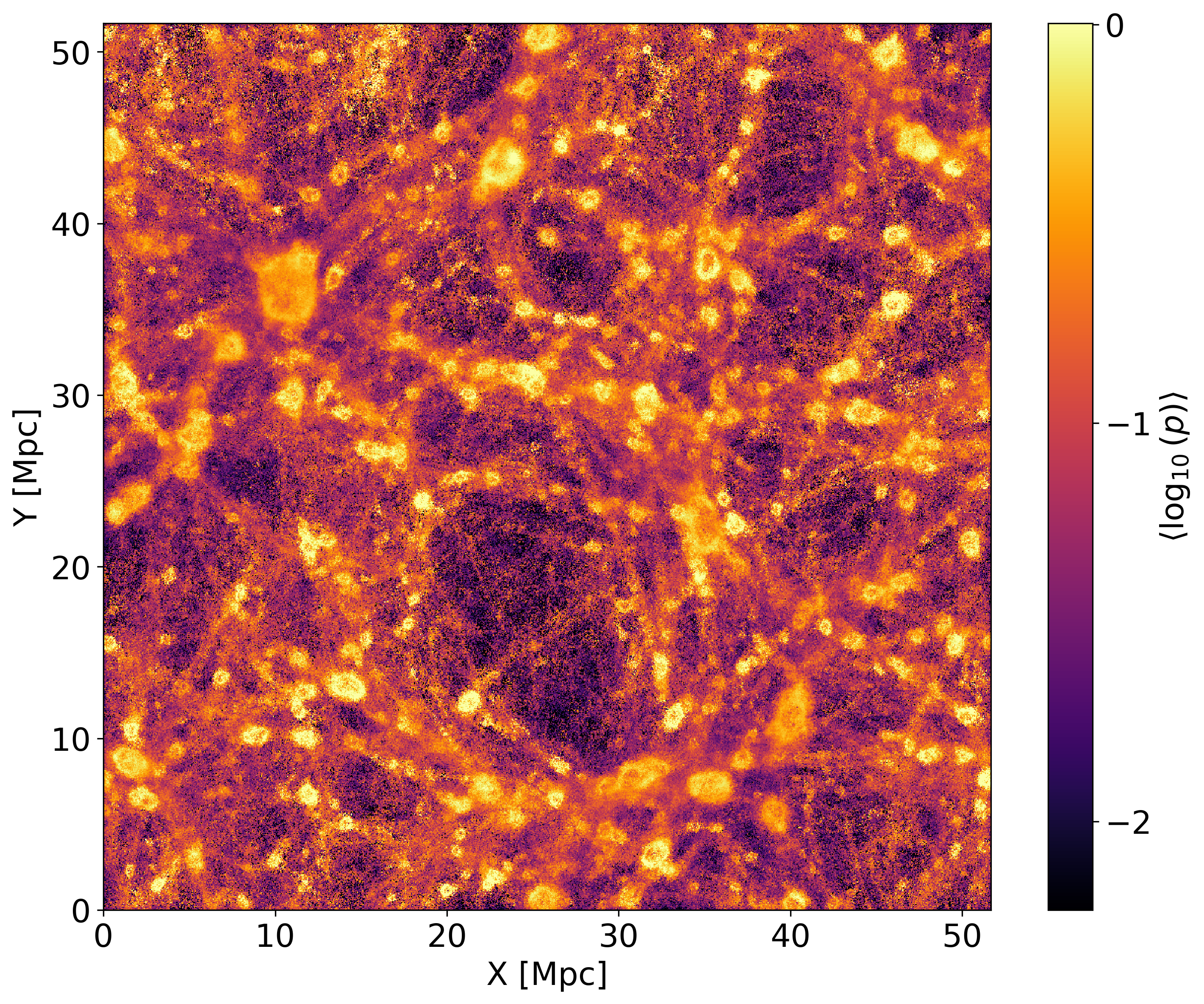}
      \caption{1-DREAM-derived pheromone distribution across the full simulation volume projected onto the $x$–$y$ plane. DM halos are binned onto a 2D grid, with each bin colored by the mean $\log_{10}(p)$ of the halos it contains. The color scale is clipped at the 1st and 99th percentiles for visualization.
      }
        \label{fig:pheromone_proj}
   \end{figure}
   
\section{Simulation Data}
\label{sec:data}

   In this work, we use the publicly available suite of cosmological simulations from the IllustrisTNG project \citep{IllustrisTNG, pillepich2019, nelson2019first}. Specifically, we focus on the TNG50-1-Dark box, a dark-matter-only (DMO) run adopting the following cosmological parameters: $\Omega_{\Lambda,0} = 0.6911$, $\Omega_{m,0} = 0.3089$, $\Omega_{b,0} = 0.0486$, $\sigma_{8} = 0.8159$, $n_s = 0.9667$, and $h = 0.6774$ \citep{planck2016}. 

   The TNG50-1-Dark box has a side length of $51.7 \, \text{Mpc}$ and contains $2160^3$ resolution elements, offering the highest mass resolution among the TNG DMO runs. The mass of a single DM particle is $m_{\text{DM}} = 5.4\times10^{5} \, M_{\odot} / h$. The simulation provides 100 snapshots (0-99), spanning redshift $z=127$ to the present day. Our analysis focuses on snapshot 99, corresponding to redshift $z=0$. The total DM particle distribution ($1.008\times10^{10}$ particles) in the simulation was processed by the TNG collaboration to generate halo catalogs using the Friends-of-Friends (FoF) algorithm \citep{davis1985} with a standard linking length parameter of $b = 0.2$, expressed as a fraction of the mean inter-particle separation. A total of 5,851,049 FoF halos are identified in the volume at $z=0$, on which our analysis is based. From this catalog, we mainly use the GroupPos, GroupVel, and GroupMass fields, corresponding to the 3D position, 3D velocity, and total FoF halo mass, respectively.

\section{Methodology}
\label{sec:methodology}

   In this section, we describe the methodology used to extract and analyze a sample of cosmic filaments embedded within the TNG50-1-Dark simulation box. Our approach is split into three parts: in Section \ref{sec:1dream_methods} we provide an overview of the 1-DREAM toolbox and its calibration for filament extraction specific to our dataset; in Section \ref{sec:classification_methods} we refine our selection of halos belonging to filaments and furthermore divide our filament sample into three density-based categories; finally, we describe our kinematic analysis in Section \ref{sec:velfield_methods}, including the construction of the cosmic velocity field within the simulation box and the derivation of its divergence/vorticity components.

    \begin{figure}
    \centering
      \includegraphics[width=\hsize]{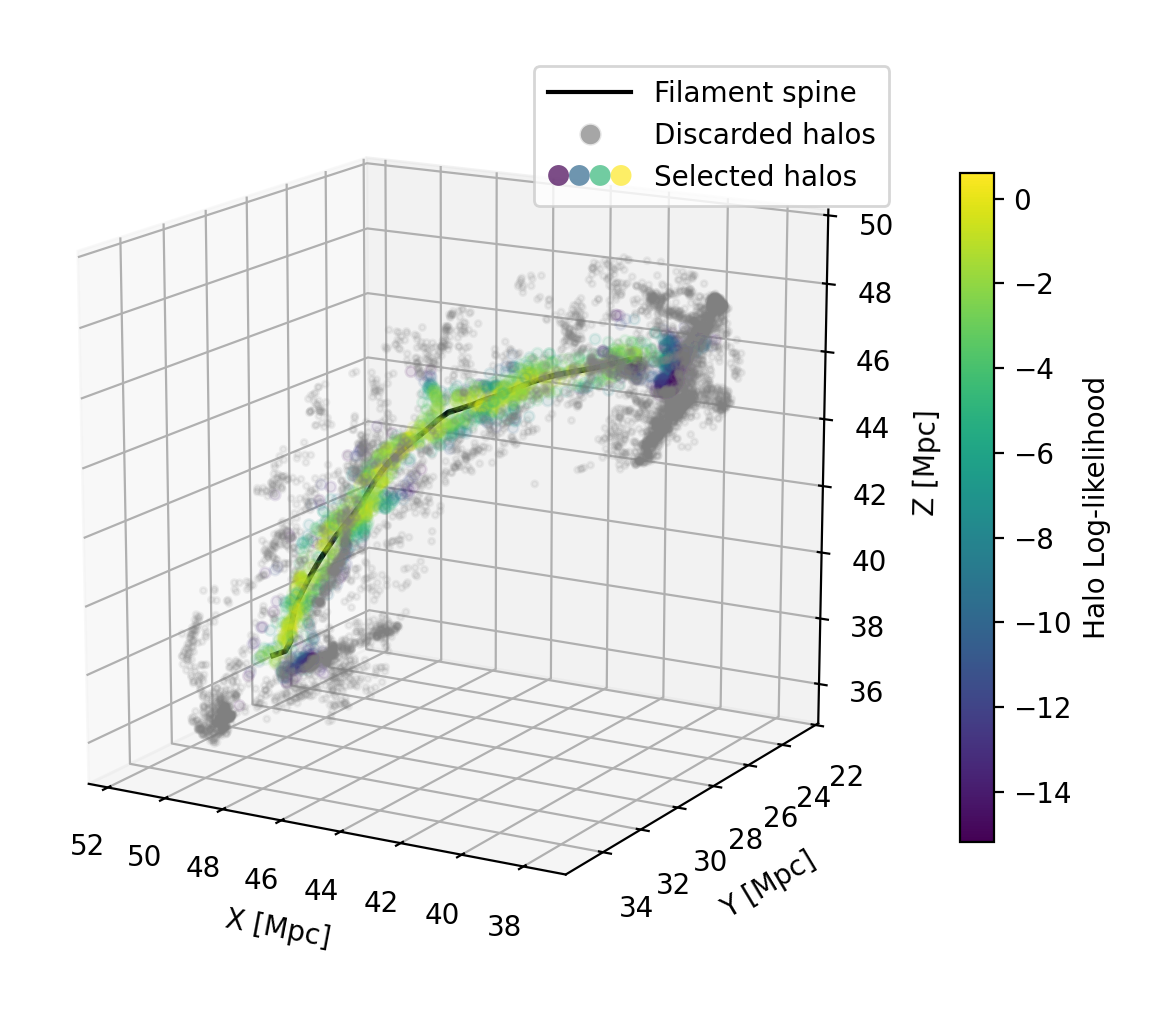}        
      \caption{Three-dimensional distribution of halos around an example filament demonstrating the refined halo selection. The filament spine is shown by the black curve, and all halos within a distance of $3\,\mathrm{Mpc}$ from the spine are included. Only halos above the likelihood threshold are colored according to the logarithm of their GMM-derived likelihood, otherwise, they are displayed in grey color.}
      \label{fig:3D_GMM}
   \end{figure}

\subsection{Filament Extraction via 1-DREAM}
\label{sec:1dream_methods}

   To identify and model filamentary structures, we employed the 1-DREAM toolbox \citep{canducci20221dream}. This framework integrates five interpretable machine learning algorithms designed to study 1-dimensional manifolds (e.g. filaments) within large point-cloud datasets. A comprehensive description of the cumulative toolbox is provided in \citet{canducci20221dream}; here, we briefly summarize the role of each algorithm, with full details on the used parameters listed in the appendix Table \ref{table:1dream_full_params}:

   \begin{itemize}
       \item The \emph{Locally Aligned Ant Technique (LAAT)} enhances the contrast between under- and over-dense regions and highlights elongated structures within the dataset using concepts from Ant Colony Optimization: a set of agents are initialized on a random walk within the data to look for the structures of interest, reinforced by an artificial quantity termed the ``pheromone" ($p$) which traces and accumulates on structures within the pointcloud. A threshold applied on the pheromone extracts the structures of interest from the remaining dataset.   
       \item \emph{Manifold Blurring Mean Shift (MBMS)} iteratively moves halos toward an estimated central axis of each identified structure in the data, further enhancing the contrast between underdense and overdense areas and highlighting the spine of cosmic structure. 
       \item \emph{Dimensionality Index (DimIndex)} assigns to each halo a dimensionality (1D, 2D, 3D) corresponding to the dimension of the structure they most likely belong to (filament, sheet, cluster, respectively). 
       \item \emph{Multi-manifold Crawling (MMCrawling)} traces the distributions of halos assigned to 1D structures to recover a catalog of filament spines represented by graph structures (nodes connected by edges), and each filament's corresponding sample of halos surrounding the spines.
       \item \emph{Stream Generative Topographic Mapping (SGTM)} utilizes the concept of constrained Gaussian Mixture Modeling (GMM) to refine the positions of the nodes tracing the filament spines and to build a probabilistic model of the surrounding halo distribution, thus providing a likelihood measure for each halo to belong to a modeled filament.     
   \end{itemize}
    
   We ran LAAT on the initial halo sample, which outputs all of the halos and their associated $p$ value, corresponding to the accumulated pheromone after all of the agents' visits. Figure~\ref{fig:pheromone_proj} displays a projection of the simulation volume following this application, with the halo distribution binned in the $x$-$y$ plane and colored by the mean $\log_{10}(p)$ in each bin. A threshold of ten times the minimum pheromone deposited in the run was applied to extract the structures of interest in the simulation. The choice of this threshold was determined by visual inspection of the halo distribution in $10\,\rm{Mpc}$-thick slices of the simulation volume, plotting halos above various pheromone deposition thresholds (ranging from 2 to 50 times the minimum $p$). The selected factor of 10 provided a compromise between effectively removing sparse, noisy regions and preserving the cosmic structures without fragmenting them. After applying this criterion, 4,261,287 halos were retained by LAAT as part of the identified structure, while 1,589,762 halos were discarded as background noise. The retained halos were then processed by MBMS to highlight the filament spines, and were subsequently partitioned by DimIndex into three structural categories shown in Figure~\ref{fig:laat_by_dimidx}, resulting in 2,932,486 halos classified as belonging to filaments. This is broadly consistent with the finding of \cite{cautun2014} that filaments account for $\sim 50\%$ of the matter in the Universe, as the number of halos assigned to filaments is approximately half of the entire halo catalog of the simulation box. 

    \begin{table*}
    \caption{Overdensity-based classification of valid cosmic filaments satisfying $L_{\mathrm{fil}} > 3.5\,\mathrm{Mpc}$.}
    \label{table:density_summary}
    \centering
    \begin{tabular}{l c c c c c}
    \hline\hline
    \noalign{\smallskip}
    Overdensity & Overdensity & N filaments & N halos & N halos & Median $\log_{10}(M_{\mathrm{h}}/M_\odot)$ \\
    Category & $\delta_{\rm{fil}}$ & ($L_{\rm {fil}}>3.5$ Mpc) & ($r=0.5$ Mpc) & ($r=3$ Mpc) & ($r=3$ Mpc) \\
    \noalign{\smallskip}
    \hline
    Low          & $\delta_{\rm{fil}} < 4$        & 703 & 431,249 & 805,351 & 7.524 \\
    Intermediate & $4 \leq \delta_{\rm{fil}} < 10$  & 224   & 589,524 & 926,149 & 7.495 \\
    High         & $\delta_{\rm{fil}} \geq 10$      & 77    & 646,981 & 1,061,385 & 7.479 \\
    \hline
    All Valid    & ---                 & 1,004 & 1,667,754 & 2,792,885 & 7.509 \\
    \hline
    \end{tabular}
\tablefoot{From left to right, we include: the category of filament overdensity, the range of overdensity over-which the category is defined, the total number of valid filaments in that category, the respective total number of halos within 0.5~Mpc of the filaments' spines, the total number of halos within 3~Mpc of the filaments' spines, and the median halo mass within each category considering a limiting radius of 3~Mpc around each filament spine. The local overdensity is defined as $\delta_{\rm{fil}} \equiv n_{\mathrm{fil}} - \bar{n} / \bar{n}$, where $\bar{n}$ is the cosmic mean halo number density. The $\delta_{\rm{fil}}$ thresholds were selected to provide a meaningful separation between diffuse, moderately overdense, and highly overdense filamentary environments and have a comparable number of halos in each category.}
    \end{table*}

    \begin{figure}
    \centering
    \includegraphics[width=\hsize]{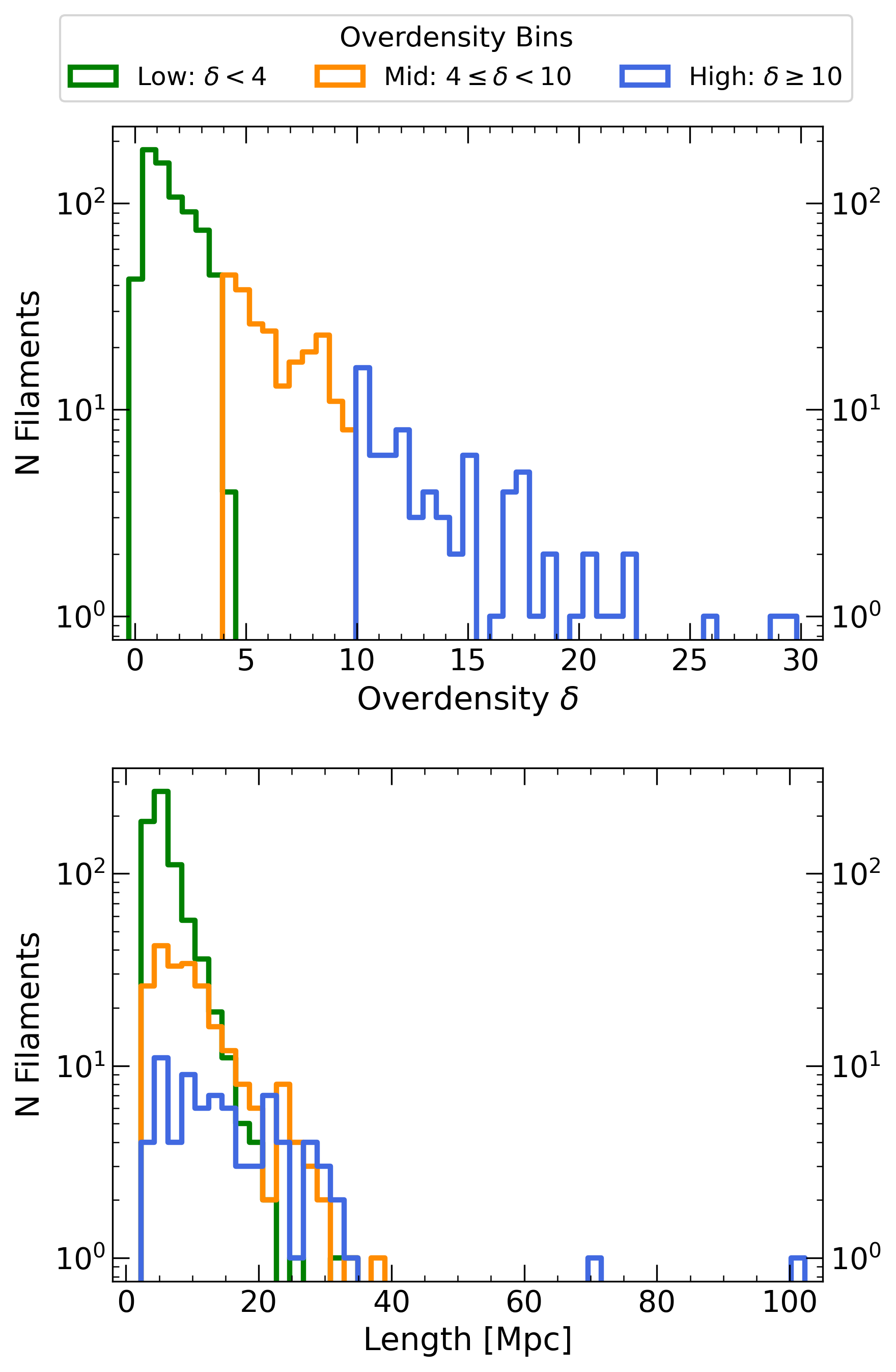}
      \caption{Overdensity (top) and length (bottom) distributions of the valid filaments sample ($L_{\rm fil}>3.5$ Mpc). Filaments are grouped by local overdensity: low  (green), intermediate (orange), and high (blue). The y-axis is shown on a logarithmic scale.}
         \label{fig:catalog_distribution}
   \end{figure}

   Tracing the one-dimensional partition of halos with MMCrawling, we identified a sample of $4,368$ preliminary filament spines and their associated halo subsets within a radius $r=0.5$ Mpc from the filament spine. We explain the selection of halos farther away from the spines later in this section. Each spine is represented as a piecewise-linear curve, composed of multiple nodes connected by edges, whose summed lengths define the total filament length. The extracted filaments were assigned unique identifiers from $1$ to $4,368$, with the first filament being the longest ($\sim100$ Mpc, see top panel of Fig.~\ref{fig:longest_filaments}). As shown in \cite{RajEtal2024}, small or undersampled filaments require a modified version of MMCrawling to be run to be certain of the validity of such filaments. To avoid such an issue, we impose a minimum length threshold on the filament catalog of  $L_{\rm fil} > 3.5\,\mathrm{Mpc}$, where the MMCrawling results for filaments that satisfy this criterion are observed to be better defined. Applying this length criterion results in a sample of 1,004 valid filaments.

   Finally, a probabilistic model of each filament is constructed by initializing a multi-variate Gaussian distribution on the position of each MMCrawling node and fitting for the relative weights and covariance matrices of the distributions using the sample of halos at a distance less than 0.5~Mpc for training. This approach keeps the positions of distributions fixed unlike the standard application of GMMs and SGTM, and was adopted to allow the modelling of both densely sampled filaments as well as sparsely sampled ones whose models using the standard SGTM approach tend to be less dependable.

   In this way, we obtain reliable spine representations of each filament in the sample as well as a likelihood measure for each halo to belong to a given modeled filament. Given a filament in the simulation, the obtained spine is used as a reference point to which we can define the Euclidean distance between the filament's axis and a given halo position, and the probabilistic model is used to search for halos at larger distances than $0.5\,\mathrm{Mpc}$ that could belong to the given filament. The latter is done by first fitting a cubic spline to the spine of the filament and using the fit to upsample the number of reference points along the spine. \citet{aragon2010} found that filament density profiles have a typical radial extent of the order of $2\,h^{-1}\,\mathrm{Mpc}$, corresponding to about $3\,\mathrm{Mpc}$ for the cosmology adopted here. More recent analyses likewise find that filament density profiles can extend to comparable distances: \citet{bahe2025} report that the mean galaxy, dark-matter, and gas overdensity profiles in EAGLE and TNG100 approach the cosmic mean within $3\,\mathrm{Mpc}$ of the filament spine, while \citet{YangEtal2025} find IllustrisTNG filament-segment radii extending to $2.75\,\mathrm{Mpc}$. 
   
   We therefore searched for candidate halos within a radius of $3\,\mathrm{Mpc}$ from the upsampled spine and then evaluated their likelihood of belonging to the filament using its trained probabilistic model. We verified the effect of the training aperture choice by testing the sensitivity of our results with different radii (see Appendix~\ref{app:gmm_training}). Different likelihood thresholds are then adopted depending on the filament density, as detailed in Section~\ref{sec:classification_methods}. This approach has thus allowed us to define a probabilistic measure for fine-tuning the filament halo sample rather than choosing halos within a fixed distance around each filament. This increases the purity of the sample as it limits the possibility of including halos that could belong to other nearby structures. An example of the GMM application and halo selection is shown in Figure~\ref{fig:3D_GMM}. Halos below the likelihood threshold are colored in grey and represent those discarded from the sample.

    \begin{figure*}
    \centering
    \includegraphics[width=\hsize-2.0cm]{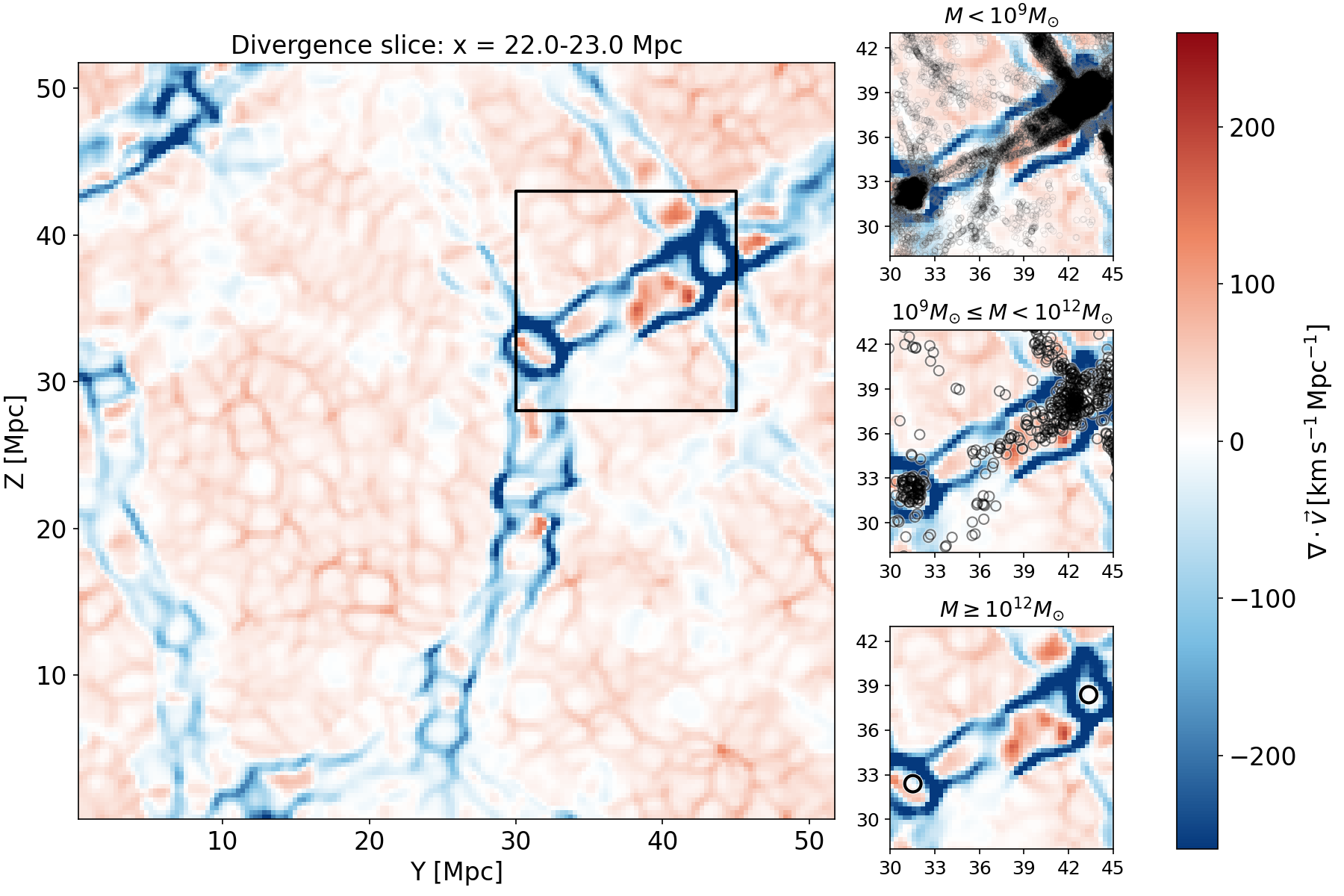}
    \par \medskip
    \includegraphics[width=\hsize-2.0cm]{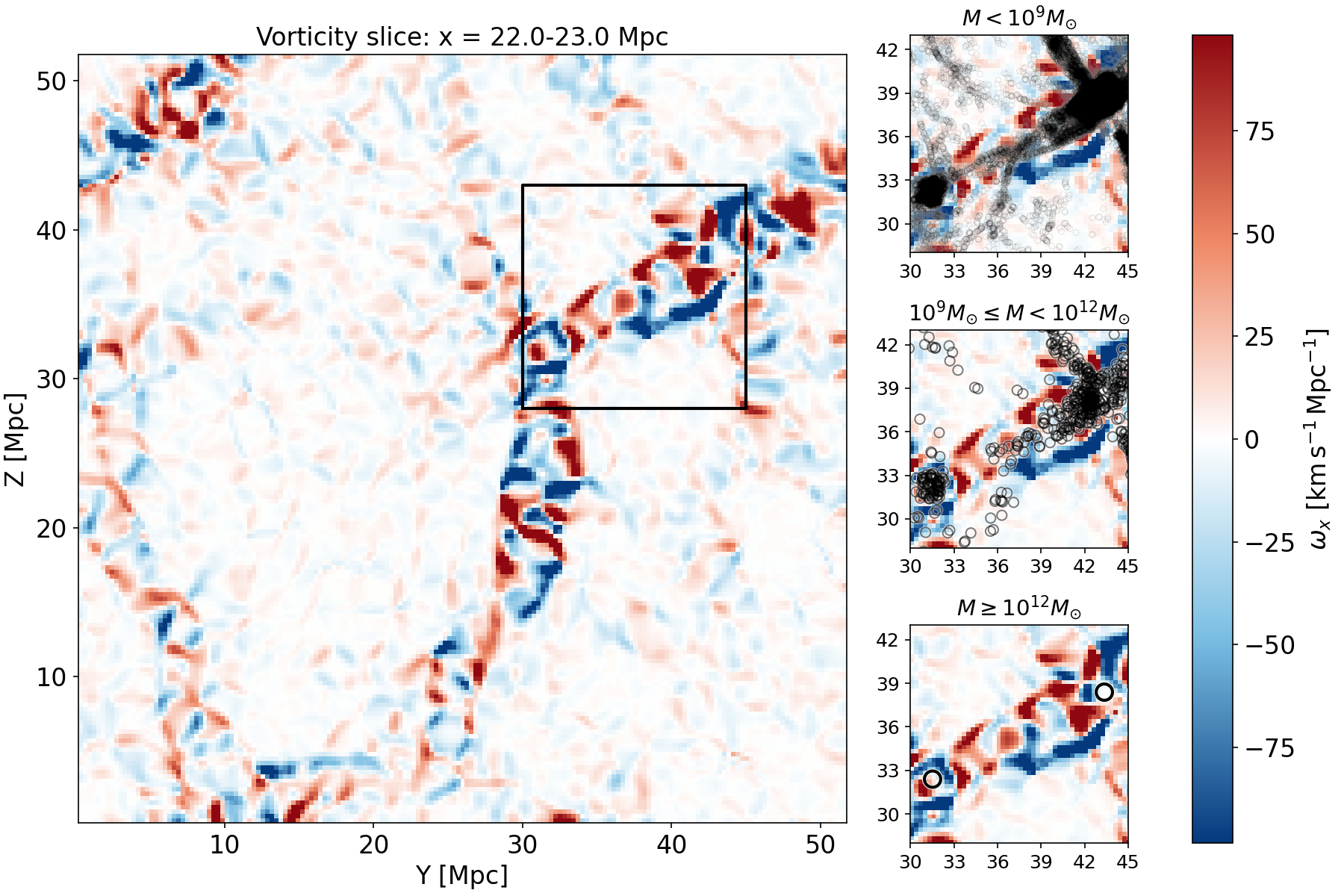}
      \caption{A 1 Mpc slice of the simulation volume in the $y$--$z$ plane at $x = 22$--$23\,\mathrm{Mpc}$. Top: velocity divergence field, $\nabla \cdot \mathbf{v}$; blue (red) regions denote sinks (sources). Bottom: $x$-component of vorticity field, $\omega_x$; blue (red) regions indicate clockwise (counterclockwise) rotation in the $y$--$z$ plane. The left panels show the full slice, while the right panels present zoom-ins of the same highlighted region (black box), overlaid with halos (circles) in three mass bins: $M < 10^{9}\,M_{\odot}$ (top), $10^{9} \leq M < 10^{12}\,M_{\odot}$ (middle), and $M \geq 10^{12}\,M_{\odot}$ (bottom). The velocity field is smoothed with a Gaussian kernel of $\sigma = 0.4\,\mathrm{Mpc}$ prior to computing spatial derivatives.}
         \label{fig:div_curl_proj}
    \end{figure*}

    \begin{figure*}
    \centering
    \includegraphics[width=\textwidth]{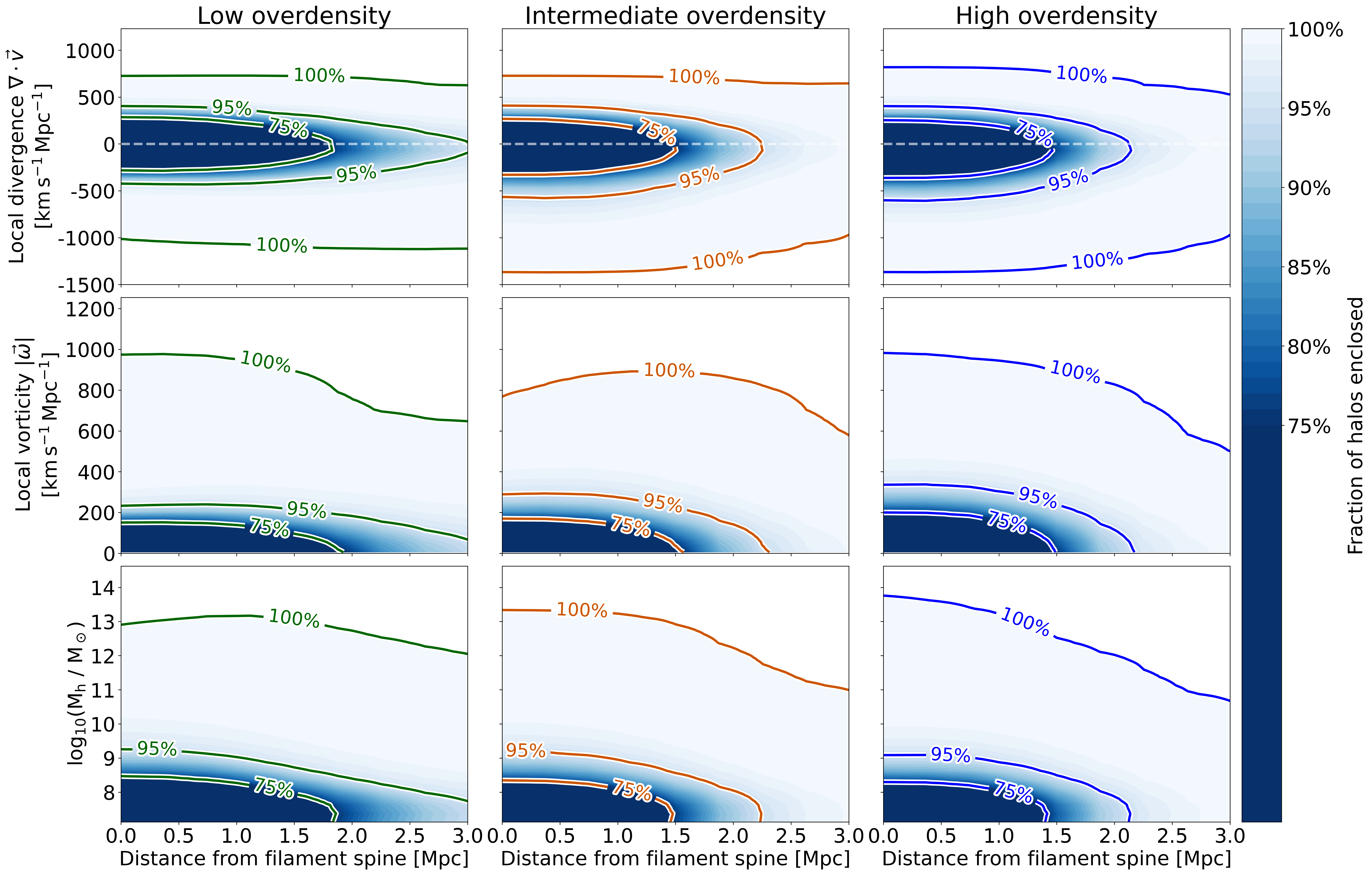}
      \caption{Halo properties as a function of distance from the filament spine, split by local overdensity environment (columns: low, mid, high) and shown for three quantities (rows: local divergence, local vorticity, and halo mass $\log_{10}(M_h/M_{\odot})$. In each panel, the background shading shows the fraction of halos enclosed within a given distance-quantity region. The colorbar is truncated at 75\%, so all regions with an enclosed fraction below this threshold appear as the same solid dark blue to ensure better contrast of the contour distribution. Contours mark the 75\%, 95\%, and 100\% enclosed-fraction levels. The white horizontal dashed line in the top row panels marks the inflow-outflow transition.}
         \label{fig:radial_plots}
    \end{figure*}

\subsection{Filament Classification}
\label{sec:classification_methods}

   Following the construction of our final filament catalog, we classify filaments within three density bins to distinguish if any trends we find between halo mass and the cosmic velocity field are present irrespective of local density. For each filament, we define a volumetric halo number density,
    \begin{equation}
    n_{\mathrm{fil}}(r) = \frac{N_{\mathrm{halos}}}{\pi r^2L_{\rm fil}},
    \end{equation}
    where $N_{\mathrm{halos}}$ is the number of halos associated with the filament and $L_{\rm fil}$ is the total filament length, calculated by summing the lengths of the edges connecting consecutive spine nodes. The effective filament volume, $V_{\mathrm{fil}}=\pi r^2L_{\rm fil}$, approximates the region surrounding the piecewise-linear spine as a cylindrical tube of constant radius $r=0.5$ Mpc, corresponding to the Euclidean distance from the filament spine. This local density was normalized by the cosmic mean halo number density,
    \begin{equation}
    \bar{n} = \frac{N_{\mathrm{tot}}}{V_{\mathrm{box}}}
    = 42.34\ \mathrm{Mpc^{-3}},
    \end{equation}
    computed from the full TNG50-1-Dark simulation box at $z=0$, where $N_{\mathrm{tot}}$ is the total number of halos in the box and $V_{\mathrm{box}}$ is the full simulation volume.
    We then define the filament overdensity:
    \begin{equation}
    \delta_{\rm{fil}} \equiv \frac{n_{\rm{fil}}(r)-\bar{n}} {\bar{n}}.
    \end{equation}

    This overdensity-based metric provides a measure of filament environment that is independent of its length and allows direct comparison across the full catalog. Using $\delta_{\rm{fil}}$, we split the 1,004 valid filaments into three overdensity categories: low, intermediate, and high overdensity filaments, as summarized in Table~\ref{table:density_summary}. Although the number of filaments in each category is variable, the total number of halos per category is comparable and sufficiently large for robust statistical analysis. This choice of classification is intentional, as all subsequent statistical analyses in Section~\ref{sec:results} are performed on the cumulative halo samples in each category irrespective of the number of filaments. In this sense, high-density filaments are treated as rare but dynamically extreme environments typically associated with strong inflows and proximity to massive clusters, while low-density filaments represent the more prevalent contribution to the cosmic web. We verified that shifting the overdensity boundaries from their fiducial values of 4 and 10 to values within 3–5 and 8–12, respectively, does not alter the qualitative trends reported in this work, although the number of filaments and halos assigned to each category change. As noted in Section~\ref{sec:1dream_methods}, the likelihood threshold applied during halo selection at $r=3$ Mpc also depends on filament overdensity (halos above the 70th, 60th, and 50th likelihood percentile are retained for low, intermediate, and high overdensity filaments, respectively), and the resulting halo counts for each category are summarized in Table~\ref{table:density_summary}.

   The top panel of Figure~\ref{fig:catalog_distribution} shows the overdensity distribution of the final filament catalog. The distribution spans a broad range, from slightly negative values ($\delta \sim0$) up to $\delta\sim30$, confirming that filaments populate a wide range of density environments. The majority of filaments reside in low and mildly overdense regions. This broad distribution is consistent with previous studies showing that the density range occupied by filaments overlaps with that of other cosmic-web environments, particularly sheets, and that density alone therefore does not uniquely determine web morphology \citep{cautun2014}. Here, however, the subsequent analysis is restricted to halos associated with filaments. The bottom panel shows the distribution of filament lengths for each overdensity category. A substantial overlap in length is observed across all three regimes, indicating that filament length is largely independent of overdensity in our sample.

\subsection{Velocity Field Construction and Calculations}
\label{sec:velfield_methods}

    The velocity field is constructed on a regular three-dimensional grid spanning the full simulation volume. All data from the initial halo catalog prior to the application of 1-DREAM is included, ensuring a globally defined field before focusing on filaments only. The box is discretized into cubic voxels with spatial resolution $\Delta x = 0.3 \,\mathrm{Mpc}$, chosen to be slightly larger than the mean inter-halo separation ($\sim 0.28\,\mathrm{Mpc}$) in order to reduce sampling noise while preserving filament-scale structure. Further grid resolutions are tested and discussed in Appendix~\ref{app:resolution}.

    Within each voxel, we compute the mass density $\rho$ and momentum density $\rho \mathbf{v}$ given the location of each halo within the grid voxels. The velocity field is then defined as the ratio of these quantities, yielding a discrete, mass-weighted velocity field that traces the dominant gravitational flows rather than being dominated by low-mass halo noise. To obtain a continuous field suitable for spatial differentiation, we apply Gaussian smoothing to both the mass and momentum fields. Assuming periodic boundary conditions (consistent with the TNG simulation volume), the smoothed velocity field is defined as
    \begin{equation}
     \mathbf{v}_{\sigma}(\mathbf{x}) =
     \frac{G_{\sigma} * \left[ \rho(\mathbf{x}) \, \mathbf{v} (\mathbf{x}) \right]}
          {G_{\sigma} * \rho(\mathbf{x})} \, ,
    \end{equation}
    where $G_\sigma$ is a 3D Gaussian kernel of width $\sigma = 0.4\,\mathrm{Mpc}$ ($\sim1.3\,\Delta x$), and $*$ denotes convolution. The smoothing scale is chosen to be slightly larger than the grid spacing while remaining smaller than characteristic TNG filament radii \citep[$\sim 1.3$-$1.5\,\mathrm{Mpc}$ at $z=0$;][]{YangEtal2025}. This corresponds to a mass-weighted coarse-graining of the discrete halo velocity field, which regularizes small-scale fluctuations and ensures that spatial derivatives remain well-defined even in multi-stream regions. 

    From the smoothed velocity field, we compute the divergence and vorticity fields via finite-difference approximations of the spatial derivatives. The divergence, $\nabla \cdot \mathbf{v}_{\sigma}$, yields a scalar field that quantifies the local rate of expansion ($\nabla \cdot \mathbf{v_{\sigma}} > 0$) or compression ($\nabla \cdot \mathbf{v_{\sigma}} < 0$), while the vorticity, $\boldsymbol{\omega} = \nabla \times \mathbf{v}_{\sigma}$, produces a vector field which describes the local rotational component of the flow. We will drop the $\sigma$ subscript from here onwards for simplicity. Spatial derivatives are computed on the voxel grid through the gradient operator, which evaluates partial derivatives of each velocity component along the three coordinate axes. The resulting voxel-level fields are mapped back to the halo catalog using a nearest-voxel assignment, where the position of each halo is digitized onto the grid, and the halo is assigned the field value of the voxel containing it. This way, all halos within the same $0.3$ Mpc voxel receive the same local divergence and vorticity values; no interpolation between neighboring voxels is applied.

\section{Results}
\label{sec:results}

    \begin{figure}[h!]
        \centering
        \includegraphics[width=\hsize]{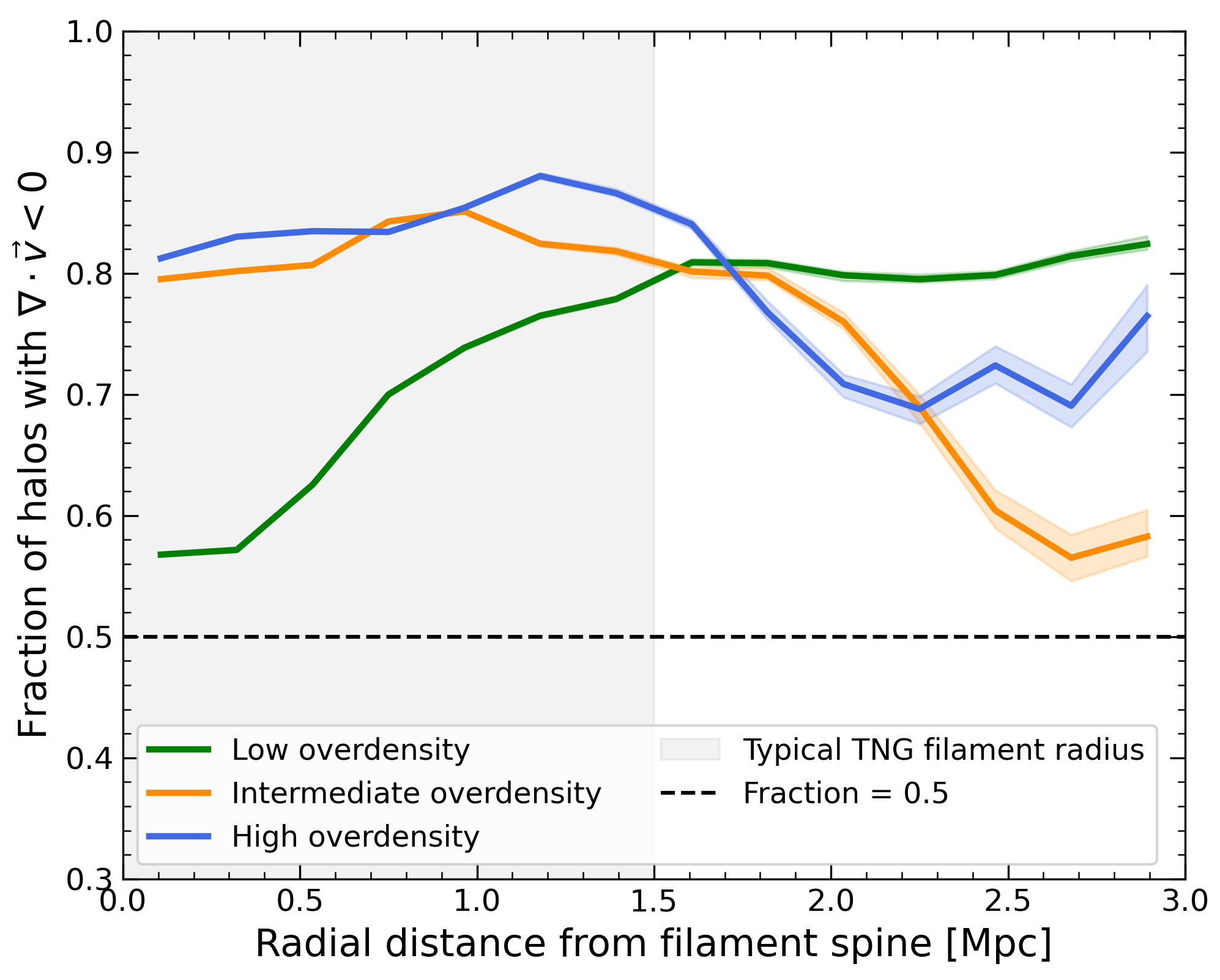}
        \caption{Fraction of halos with negative local velocity field divergence ($\nabla \cdot \mathbf{v} < 0$) as a function of radial distance from the filament spine for low (green), intermediate (orange), and high (blue) overdensity filaments. Uncertainty margins are calculated through bootstrapping from the halo sample per distance bin. The shaded area corresponds to typical filament widths in the IllustrisTNG simulations \citep{YangEtal2025}.}
        \label{fig:div_dist}
    \end{figure}

    By inspecting slices of the simulated volume, we find that the global topology of the velocity field is highly structured, as evidenced by the spatial distribution of the divergence field shown in the top panel of Figure~\ref{fig:div_curl_proj}. Regions of converging flow ($\nabla \cdot \mathbf{v} < 0$) are strongly correlated with the cosmic web's skeleton, tracing overdense structures and peaking at the nodes of the network. In contrast, diverging flow ($\nabla \cdot \mathbf{v} > 0$) dominates the large, underdense volumes, representing the evacuation of matter from cosmic voids. Additionally, we observe some positive divergence in the innermost regions of collapsed structures, where relaxation takes place \citep{zhu2017} and the infall slows down. This spatial configuration is consistent with the Zel’dovich approximation, where filaments and clusters act as the primary sinks for the tidal field and thus the velocity field. Notably, after inspecting multiple slices, we find that halos with masses $\geq 10^{12} M_\odot$ are preferentially located in low absolute divergence regions.

    Complementing the divergence field, the bottom panel of Figure~\ref{fig:div_curl_proj} reveals the projected vorticity ($\omega_x$), which exhibits a more localized and fragmented small scale structure. While the divergence field displays broad, continuous regions of convergence, the vorticity field is characterized by high-magnitude, intermittent fluctuations of positive and negative vorticity, corresponding to counterclockwise and clockwise rotation in the projected plane, respectively. Vorticity amplitudes are generally lower in underdense regions and become stronger toward overdense cosmic-web structures, where more complex streams appear. This suggests that rotational flows and non-linear processes such as shell-crossing and multi-streaming are more prominent in denser cosmic-web environments. We also observe that halos with masses $\geq 10^{12} M_\odot$ frequently inhabit the boundaries of these high-vorticity patches, sitting in transition zones between opposing rotational flows, where the projected vorticity component approaches zero.

    \begin{figure*}
        \includegraphics[width=\textwidth]{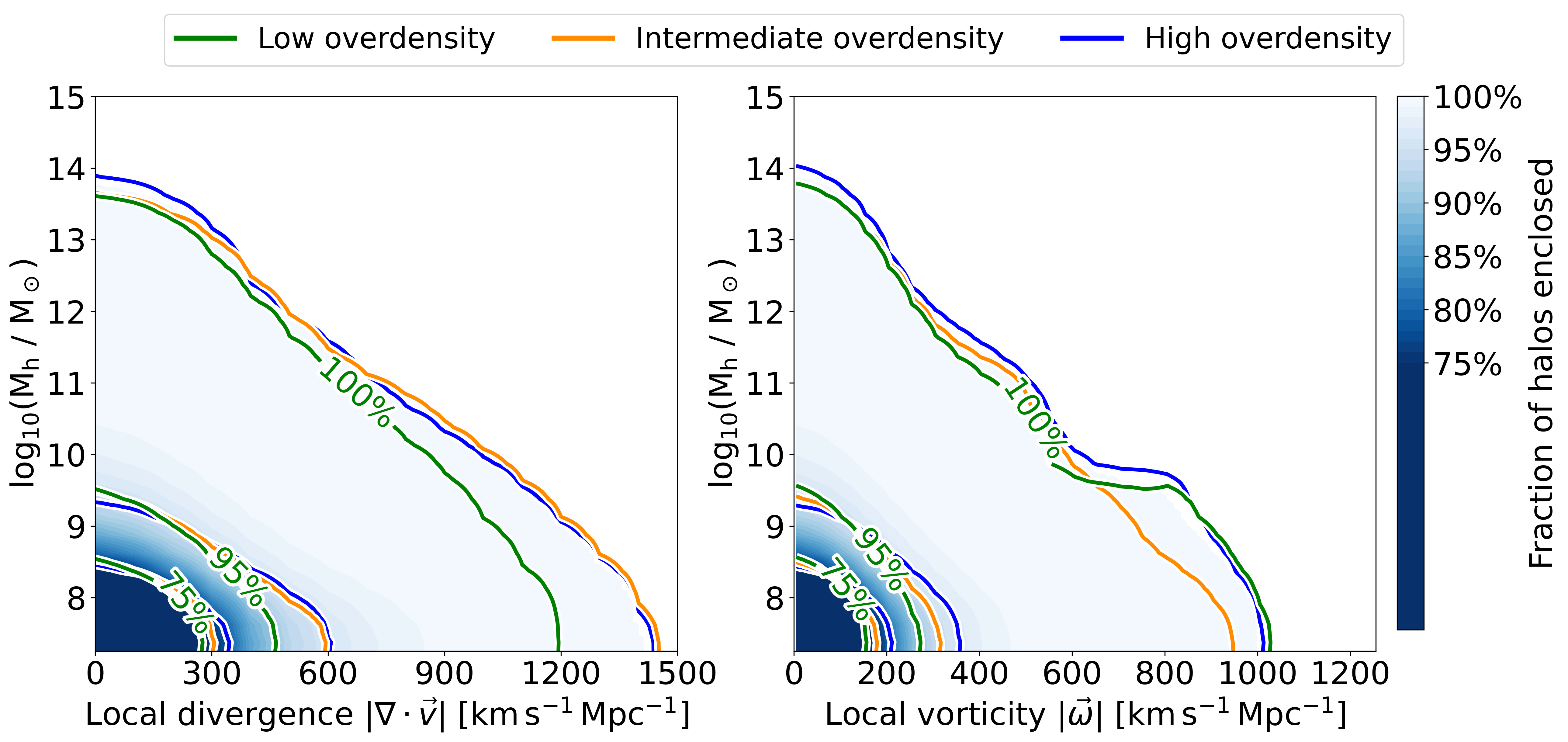}
        \caption{Halo mass, $log_{10}(M_h/M_{\odot})$, as a function of local divergence in absolute value (left) and local vorticity (right), for halos in high- (blue), mid- (orange), and low- (green) overdensity environments. Shading shows the fraction of halos enclosed for the combined sample across all three environments, with the colorbar truncated at 75\% (darker blue regions below this threshold appear as a single solid color). Colored contours mark the 75\%, 95\%, and 100\% enclosed-fraction levels for each environment individually.}
        \label{fig:mass_curl_div}
    \end{figure*}

    \subsection{Radial Distribution}
    \label{subsec:radial_dist}
    
    Focusing the analysis specifically onto our valid filaments sample, in Figure~\ref{fig:radial_plots}, we combine the halo population within each defined overdensity category and examine how three halo properties: local divergence, local vorticity, and halo mass $\log_{10}(M_h/M_{\odot})$, vary as a function of radial distance ($r$) from the filament spines. The figure shows cumulative halo fraction contour plots of the distribution of dark matter halo properties as a function of distance from the filament spine. For each of the three overdensity environments (low, mid, and high), we construct two-dimensional histograms of halo mass, local vorticity, and local divergence against distance from the filament spine using fixed bin edges for each quantity. For visual clarity, the resulting histograms are smoothed with a Gaussian filter ($\sigma = 1.5$ bins) and interpolated by a factor of 7. Each pixel in the smoothed distribution is then assigned a cumulative probability value representing the fraction of halos enclosed within that contour level, allowing us to draw iso-probability contours at 75\%, 95\%, and 99.999\% $\approx 100 \%$ enclosure fractions. To increase the color contrast and aid visibility of the outer contours, we use a blue color scale truncated to show only the region between the 75\% and 100\% levels, masking the densest central region. The columns of  Figure~\ref{fig:radial_plots} correspond to the filament overdensity categories, while the rows correspond to the mentioned halo properties. 
    
    The top-row panels show the radial evolution of $\nabla \cdot \mathbf{v}$, confirming that converging flows dominate at all distances, with the bulk of the distribution concentrated at negative divergence values at all values of $r$. Across all overdensity bins, the strongest convergence is not found at the filaments' very center (consistent with \citealt{zhu2017} and the associated infall slowdown). The divergence distribution broadens over the first $\sim 1$ Mpc before narrowing again toward larger radii. This effect is most gradual in low $\delta_\mathrm{fil}$ filaments, where the distribution is flatter with distance, suggesting a more extended infall pattern in less dense environments. When all radial bins are combined in each overdensity category, 65.24\% of halos in low $\delta_\mathrm{fil}$ filaments reside in regions of converging flow, compared with 80.97\% and 83.10\% in intermediate and high $\delta_\mathrm{fil}$ filaments, respectively. Figure~\ref{fig:div_dist} quantifies this trend more directly, showing the fraction of halos in converging flow ($\nabla \cdot \mathbf{v} < 0$) as a function of $r$. For each distance bin, the fraction uncertainties are estimated via bootstrap resampling (100 realisations), with the shaded bands indicating the 16th–84th percentile interval. In intermediate- and high-overdensity filaments, this fraction peaks at $\sim0.9$ and $\sim1.2\,\mathrm{Mpc}$ respectively before declining at larger radii. We observe that this peak occurs at the distance matching the expected widths found in \citet{YangEtal2025} for filaments within the IllustrisTNG simulations at $z=0$ (1-1.5~Mpc). The outer bound of this width range is shaded in grey in Figure~\ref{fig:div_dist}. As for low-overdensity filaments, instead of a clear peak, we observe a gradual increase in the fraction of inflowing halos as we move farther away from the filament spine, which then plateaus at the distance matching the typical filament widths. This result complements what we see in the top panel of Figure~\ref{fig:div_curl_proj}, confirming that the largest magnitudes of velocity divergence occur at the edges of cosmic filaments.

    \begin{figure*}[t]
        \centering
        \includegraphics[width=\linewidth]{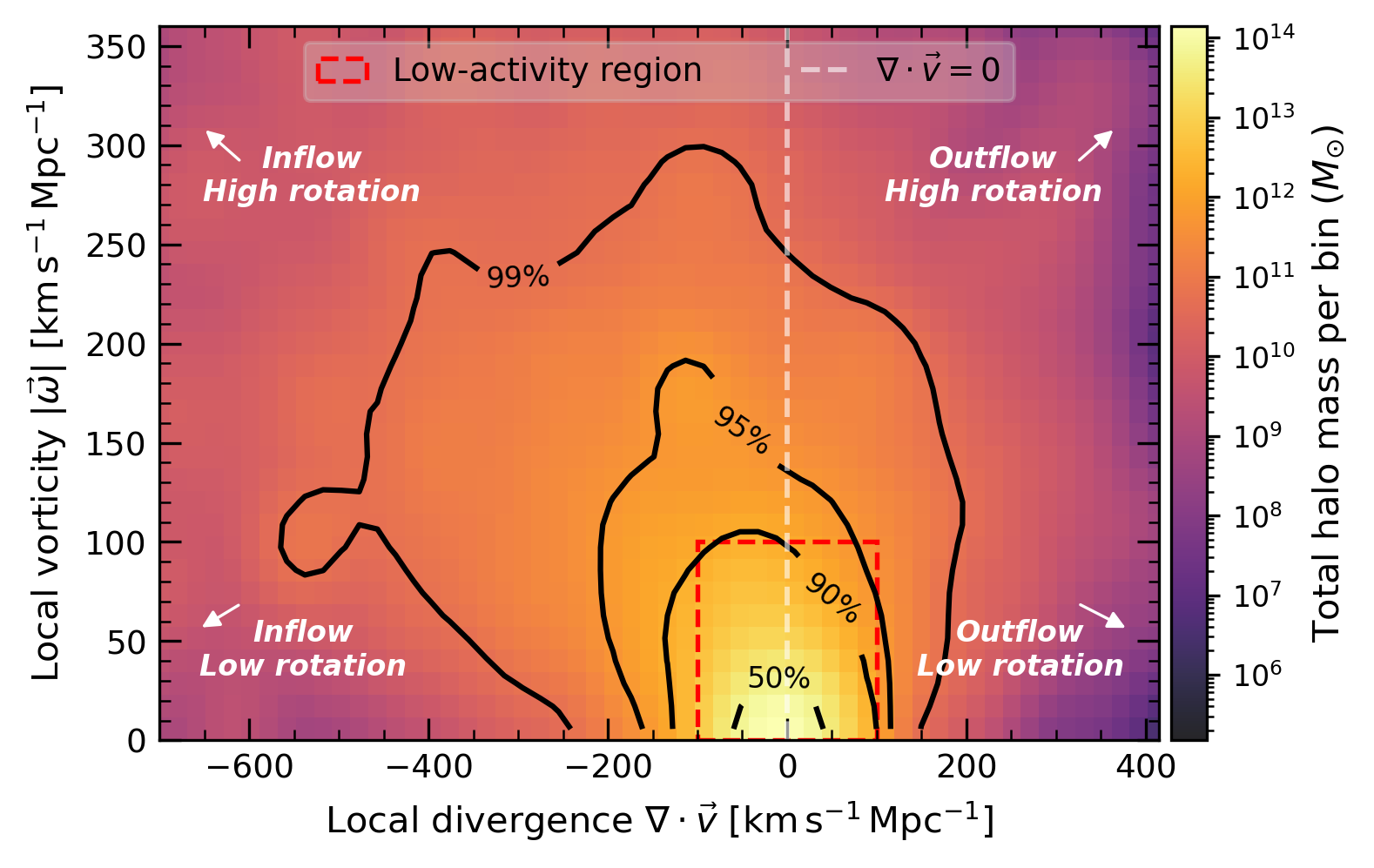}
        \caption{Smoothed total halo mass per bin in the local divergence–vorticity plane, for all halos associated with valid filaments. The colorbar shows the total halo mass per bin summed over the full sample. Black solid contours enclose 50\%, 90\%, 95\%, and 99\% of the total halo mass. Quadrant labels and arrows indicate the four combinations of flow (inflow/outflow) and vorticity (high/low rotation). The white vertical dashed line marks the inflow-outflow transition, and the red dashed box encloses the low-activity region, where $|\nabla\cdot\mathbf{v}| < 100\,\mathrm{km}\,\mathrm{s}^{-1}\,\mathrm{Mpc}^{-1}$ and $|\boldsymbol{\omega}| < 100\,\mathrm{km}\,\mathrm{s}^{-1}\,\mathrm{Mpc}^{-1}$.}
        \label{fig:money_plot}
    \end{figure*}

    Similarly, the local vorticity $|\tilde{\omega}|$ distribution reaches higher values over the inner regions of the filaments' cross sections across all overdensity bins. The strongest high-vorticity tail occurs at intermediate radii ($r \sim 0.5$--$1\,\mathrm{Mpc}$), before narrowing steadily toward the outskirts. This radial behavior suggests that vortical motions are also enhanced in the intermediate regions surrounding the filament core, where multi-stream interactions are expected to be most prominent \citep{laigle2015}. The enclosed-fraction contours in the middle-rows of Figure~\ref{fig:radial_plots} capture this trend primarily through the declining upper envelopes of the vorticity distribution beyond the inner $\sim 1$ Mpc. The effect becomes increasingly pronounced with filament overdensity, with the bulk of the distribution in high-$\delta_{\rm fil}$ filaments extending to larger vorticity amplitudes than their lower-density counterparts, which also display flatter contours.

    The bottom-row panels show clear radial stratification in halo mass, with the 75\% and 95\% contours shifting toward lower masses at increasing distance from the spine. The 100\% enclosed-fraction contour shows that the highest-mass objects are increasingly confined toward smaller radii in denser filaments. As overdensity grows, we also observe a decrease in radial spread for the most massive halos ($\geq 10^{12} M_\odot$). This trend is consistent with enhanced dynamical evolution and mass assembly in denser filamentary environments, where higher interaction rates and merger activity contribute to the buildup of massive halos near the filament spine. 
    
    The spatial segregation observed across all panels of Figure~\ref{fig:radial_plots} suggests that the inner filament regions not only concentrate mass but also correspond to dynamically distinct environments, motivating a direct comparison between halo mass and local kinematic properties.

    \subsection{Connections between halo mass and the velocity field}

    We further examine the relationship between halo mass and the local flow properties by plotting, for all three overdensity environments combined, the joint distribution of halo mass against local vorticity and absolute divergence. Similar to Figure~\ref{fig:radial_plots}, the color fill represents the cumulative halo fraction distribution this time of the combined sample, while the iso-probability contours at 75\%, 95\%, and 100\% are drawn separately for each overdensity environment.
    We find that while lower-mass halos lie in regions that span a wide range of kinematic states, the most massive halos in our sample ($M \geq 10^{12}\,M_\odot$) are strongly concentrated in regions of low absolute divergence and low vorticity. In the left panel of Figure~\ref{fig:mass_curl_div}, the 75\% and 95\% enclosed-fraction contours show that the halo-mass distribution narrows with increasing local absolute divergence $|\nabla \cdot \mathbf{v}|$, with the highest masses concentrated toward low absolute values across all three filament overdensity categories. The same behavior is seen for vorticity in the right panel, where the bulk of the halo population shifts toward lower masses as $|\tilde{\omega}|$ increases. Quantitatively, out of the 266 halos with $M \geq 10^{12}\,M_\odot$, 264 halos ($99.25\%$) have $|\nabla \cdot \mathbf{v}| < 100\,\mathrm{km}\,\mathrm{s}^{-1}\,\mathrm{Mpc}^{-1}$ while 263 halos ($98.87\%$) reside in $|\tilde{\omega}| < 100\,\mathrm{km}\,\mathrm{s}^{-1}\,\mathrm{Mpc}^{-1}$, and none exceed $200\,\mathrm{km}\,\mathrm{s}^{-1}\,\mathrm{Mpc}^{-1}$ in either quantity. Even under a more restrictive threshold of $50\,\mathrm{km}\,\mathrm{s}^{-1}\,\mathrm{Mpc}^{-1}$, $97\%$ and $96.24\%$ of massive halos remain within low absolute divergence and low vorticity, respectively. The apparent extension of the outer $100\%$ enclosed-fraction contours toward larger divergence and vorticity values is driven by the small number of objects occupying the distribution tails and is further broadened by the smoothing applied for visualization. The $75\%$ and $95\%$ contours thus provide a more representative description of the bulk halo population. 
    
    These trends indicate that extreme kinematic conditions, characterized by large absolute velocity divergence or strong rotational flows, are associated with the absence of massive halos. Our results demonstrate that halo mass is tightly coupled to the local kinematic state of the cosmic flow, with the most massive systems occupying a limited region of low absolute divergence, low vorticity environments, and preferentially residing near the filament spine, where these dynamically calmer conditions are most prevalent. The fact that such trends are consistent among all filament overdensity categories is further discussed in Section~\ref{sec:discussion}.

   To further visualize the results of divergence and vorticity jointly rather than as two independent projections, Figure~\ref{fig:money_plot} shows the total halo mass per bin in the divergence–vorticity plane for all halos within our filament sample. The contours enclosing $50\%$, $90\%$, $95\%$, and $99\%$ of the summed halo mass per bin confirm that the mass budget is overwhelmingly concentrated in a compact region of low absolute divergence and low vorticity, nearly centered on the inflow/outflow transition. We will refer to this zone where both $|\nabla\cdot\mathbf{v}| < 100\,\mathrm{km}\,\mathrm{s}^{-1}\,\mathrm{Mpc}^{-1}$ and $|\boldsymbol{\omega}| < 100\,\mathrm{km}\,\mathrm{s}^{-1}\,\mathrm{Mpc}^{-1}$ as the 'low-activity region'. Notably, all four contours are offset toward negative divergence rather than being symmetric about $\nabla\cdot\mathbf{v}=0$. This asymmetry indicates that converging flows are favored over diverging ones in the regions where most of the halo mass resides, consistent with the converging-flow fractions reported in Section \ref{subsec:radial_dist}. Figure~\ref{fig:money_plot} demonstrates that the mass segregation reported in Figure~\ref{fig:mass_curl_div} is not an artifact of considering divergence and vorticity separately, but reflects a genuine joint constraint: the most mass-rich parts of the halo population occupy regions that are simultaneously kinematically calm in both quantities (with a preference for weakly converging over weakly diverging flow), rather than merely calm in one axis while unconstrained in the other.

\section{Discussion}
\label{sec:discussion}
        
    Our results demonstrated in Figures~\ref{fig:div_curl_proj} to \ref{fig:money_plot}  show that massive halos preferentially occupy regions of low vorticity and low absolute velocity divergence, populating the inner to central regions of filaments, while lower-mass halos populate the full range of dynamical states. This behavior persists across all of the filament overdensity categories considered in this work, indicating that the velocity-gradient field carries information beyond a purely density-based description of the cosmic web, and that the dynamical state of filaments is connected to the halo mass assembly process. 

    \begin{figure}
        \centering
        \includegraphics[width=\hsize]{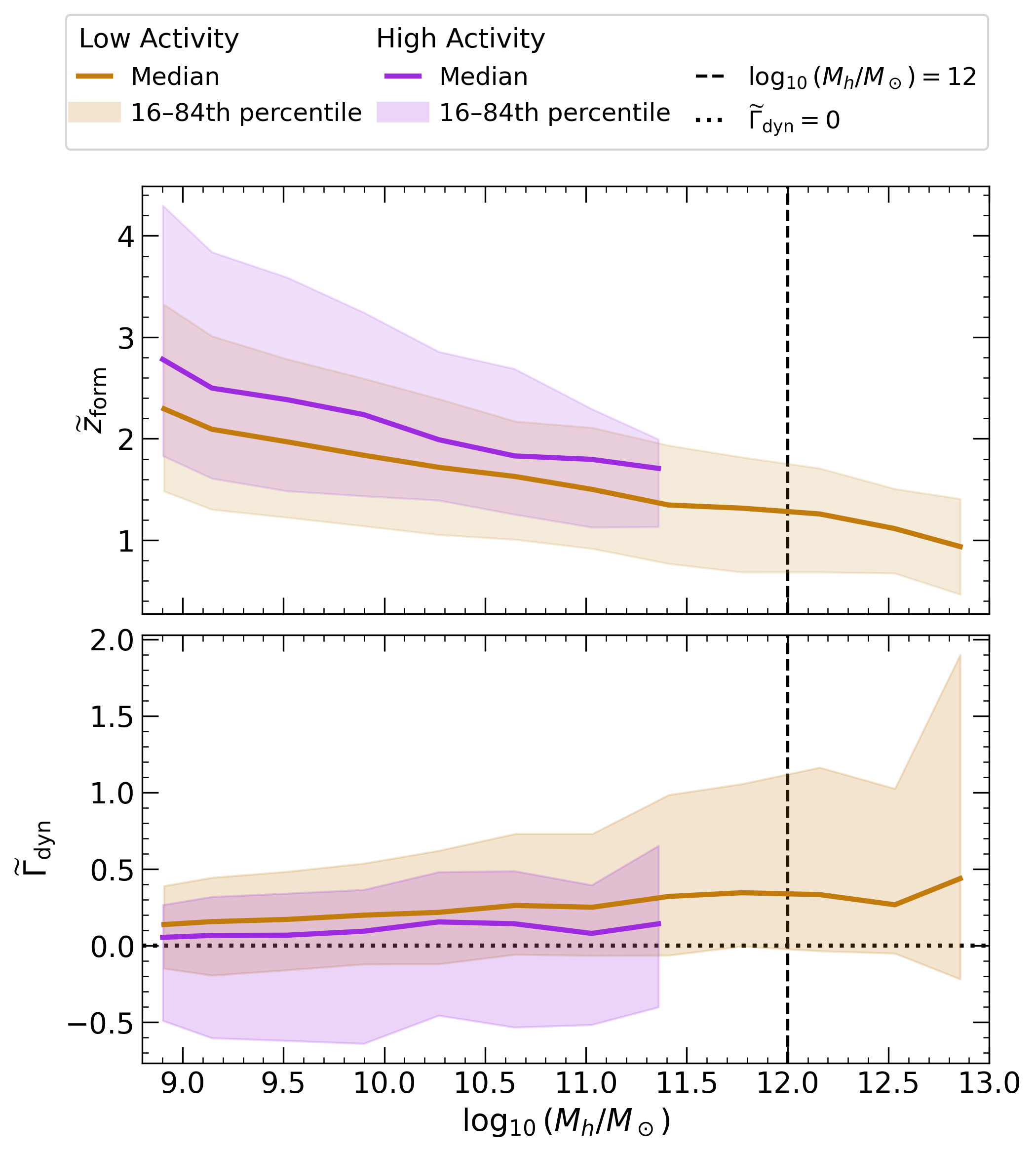}
        \caption{Median halo formation redshift ($\tilde{z}_{\rm form}$, top) and dimensionless logarithmic mass accretion rate ($\widetilde{\Gamma}_{\rm dyn}$, bottom) split by local dynamical activity and as a function of halo mass. Only filament halos with masses $> 10^{8.8}\,M_{\odot}$ are considered. Low-activity halos (gold) satisfy both $|\nabla\cdot\mathbf{v}| < 100\,\mathrm{km}\,\mathrm{s}^{-1}\,\mathrm{Mpc}^{-1}$ and $|\boldsymbol{\omega}| < 100\,\mathrm{km}\,\mathrm{s}^{-1}\,\mathrm{Mpc}^{-1}$, while high-activity halos (purple) satisfy both $|\nabla\cdot\mathbf{v}| \geq 100\,\mathrm{km}\,\mathrm{s}^{-1}\,\mathrm{Mpc}^{-1}$ and $|\boldsymbol{\omega}| \geq 100\,\mathrm{km}\,\mathrm{s}^{-1}\,\mathrm{Mpc}^{-1}$. Solid lines show the respective medians in bins of $\log_{10}(M_h/M_\odot)$; shaded bands show the 16th–84th percentile range. Bins containing fewer than 20 halos are omitted. The vertical dashed line marks $\log_{10}(M_h/M_\odot)=12$ and the horizontal dotted line in the bottom panel marks $\widetilde{\Gamma}_{\rm dyn}=0$.}
        \label{fig:mass_assembly}
    \end{figure}

        \begin{figure*}
    \centering
        \includegraphics[width=\textwidth]{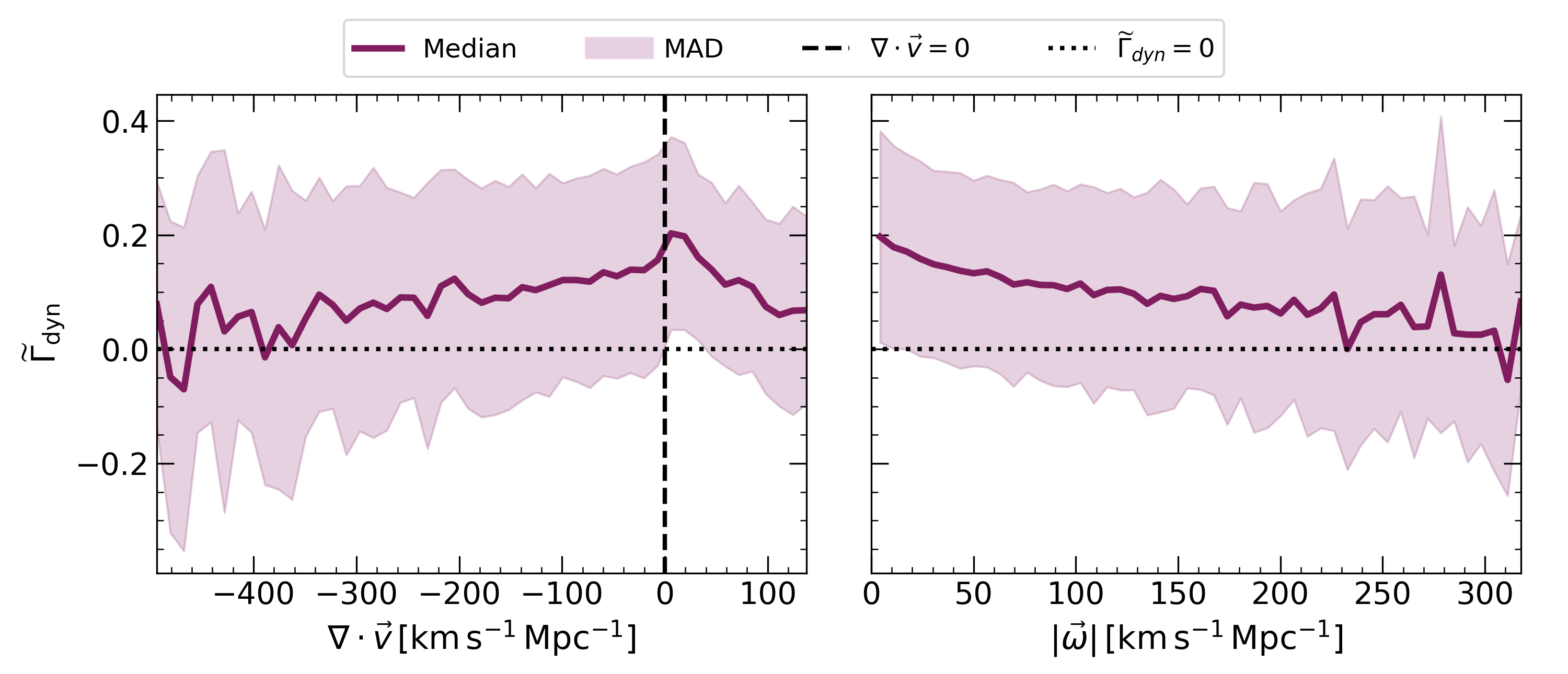}
        \caption{The median mass accretion rate per bin as a function of local velocity divergence (left) and local vorticity magnitude (right), for filament halos. In both panels, solid lines indicate the median $\Gamma_{\rm dyn}$ value within each bin, while the shaded regions represent the median absolute deviation (MAD). The vertical dashed line in the left panel marks the inflow-outflow transition.}
      \label{fig:mar_dyn_trends}
    \end{figure*}
    
    Our work connects to several results in the literature. The positive divergence observed in the innermost filament regions is consistent with \citet{zhu2017}, who showed that the innermost cores of collapsed structures undergo relaxation, causing infalling material to decelerate and producing locally diverging flow near the center. The enhancement of vorticity at intermediate radial distances from the filament spine ($r \sim 0.5$--$1\,\mathrm{Mpc}$) is in agreement with \citet{laigle2015}, who demonstrated that vorticity is generated in the mildly nonlinear regime through shell-crossing and multi-streaming, and is therefore preferentially found in the transition zones surrounding filament cores rather than in their innermost regions. More directly, \citet{etezad2025} have shown that incorporating velocity-field information at early redshifts improves the prediction of the present-day halo mass function, indicating that the density field alone does not fully capture the relevant information for halo assembly. 

    Our results described in Section~\ref{sec:results} extend this picture by demonstrating that the velocity field could have a possible regulatory role in mass-build up within halos of the large-scale structure by defining a regime of low dynamic activity that is associated with enhanced halo growth. Despite the cosmic web being highly dynamic, the most massive halos are rarely found in regions of extreme compression or extreme expansion, occupying instead kinematically quiet to moderate environments near the transition between inflow and outflow regimes. Similarly, strongly vortical environments appear to inhibit coherent large-scale accretion onto massive systems, possibly due to disrupting the smooth accretion flow required for sustained halo growth. Moreover, corroborating the findings of \citet{etezad2025}, our work suggests that mass assembly is not only explained by the density of cosmic filaments, since the trends persist across all three overdensity regimes.

    Given the results discussed above, massive halos residing in these preferred kinematic environments should exhibit distinct assembly histories. To investigate this possibility, we examine two supplementary quantities from the IllustrisTNG halo structure catalog of \citet{anbajagane2022}. Specifically, we consider the halo formation scale factor $a_{\rm form}$ (defined as the epoch at which the main progenitor halo first attained half of its present-day mass) and the dynamical mass accretion rate $\Gamma_{\rm dyn}$. Both quantities are derived from halo assembly histories traced through the SUBLINK merger trees \citep{rodriguezgomez2015}. We convert the formation scale factor into a formation redshift through $z_{\rm form}=\frac{1-a_{\rm form}}{a_{\rm form}}$. The recent mass growth is quantified using the dimensionless logarithmic accretion rate:
    \begin{equation}
        \Gamma_{\rm dyn} = \frac{\ln{M(a_{\rm{now}})}-\ln{M(a_{\rm{dyn}})}}{\ln{a_{\rm{now}}}-\ln{a_{\rm{dyn}}}},
    \end{equation}
    measured over one dynamical time as defined in \citet{diemer2017}. While the formation redshift probes when halos assembled half of their present-day mass, the mass accretion rate provides complementary information about their recent growth activity. Positive values ($\Gamma_{\rm dyn} > 0$) indicate net mass growth through accretion or mergers; values near zero indicate little net mass evolution over the past dynamical time; and negative values ($\Gamma_{\rm dyn} < 0$) indicate net mass loss, likely through tidal stripping or disruptive environmental interactions.

    Figure~\ref{fig:mass_assembly} shows both median $z_{\rm form}$ and median $\Gamma_{\rm dyn}$ as a function of halo mass, split into low- and high-dynamical-activity subsets, for filament halos with $M> 10^{8.8}\,M_{\odot}$. Halos residing in low-activity regions ($|\nabla\cdot\mathbf{v}|$, $|\boldsymbol{\omega}| < 100\,\mathrm{km}\,\mathrm{s}^{-1}\,\mathrm{Mpc}^{-1}$) exhibit lower formation redshifts and higher recent accretion rates than halos of comparable mass in the dynamically active subset. Both subsets broadly display the expected decrease of $z_{\rm form}$ with halo mass, together with a tendency for $\Gamma_{\rm dyn}$ to increase toward larger masses. This is consistent with the hierarchical picture in which massive halos assemble a substantial fraction of their mass at relatively late times while continuing to grow through mergers and anisotropic accretion along filamentary structures \citep{anbajagane2022}. Notably, the high-activity subset no longer satisfies our minimum bin-count threshold above $\log_{10}(M_h/M_\odot)\approx11.4$, and its median is not shown beyond this point. This truncation is a direct consequence of the mass segregation reported in Figure~\ref{fig:mass_curl_div}, since halos above this mass are essentially absent from the high-activity subset, so too few remain to constrain a reliable median.
    
    While the median $\Gamma_{\rm dyn}$ remains positive for both dynamical regimes in the lower panel of Figure~\ref{fig:mass_assembly}, the 16th–84th percentile band is asymmetric for both populations, although in opposite directions. The high-activity population is skewed toward negative values, with its lower percentile extending well below the median at all masses, indicating a stronger tail toward net mass loss. Conversely, for the low-activity population, we see its upper percentile extends increasingly far above the median toward higher halo mass, reflecting episodes of strong accretion that are largely absent from the high-activity subset. These trends suggest that kinematically calm regions not only tend to sustain steadier growth but also allow occasional strong accretion events that build up the most massive halos, whereas strongly rotational or strongly compressive/expanding flows are less conducive to efficient halo assembly despite often containing systems that have formed earlier.

    This picture is further reinforced by the trends shown in Figure~\ref{fig:mar_dyn_trends}. The median $\Gamma_{\rm dyn}$ does not peak in the strongest converging regions, but rather near the transition regime around $\nabla \cdot \mathbf{v} \approx 0$, and decreases systematically with increasing vorticity magnitude, supporting the interpretation that kinematically calm environments are associated with more efficient recent accretion and thus with the preferential presence of the most massive halos. We note, however, that kinematically calm environments are not universally conducive to growth. Even near $\nabla\cdot\mathbf{v} \approx 0$ and low vorticity, a population of halos with negative $\Gamma_{\rm dyn}$ is present, indicative of mass loss potentially driven by interactions with neighboring massive structures. The median trends reported here thus reflect the dominant behavior of the population rather than a strict threshold, and individual halos may deviate from the observed trend.

    This dependence on the kinematic environment is a natural target for observational follow-up. PV surveys such as CosmicFlows \citep{TullyEtal2023} have already enabled reconstructions of the local velocity field and the identification of flow-defined structures such as basins of attraction \citep{courtois2013,tully2014,courtois2023}. Ongoing and future surveys such as \textit{DESI-PV} \citep{desipv2023} and \textit{4HS} \citep{4most2023} will extend these measurements to much larger PV samples. The reconstructed velocity fields from such surveys encode the divergence and vorticity structure studied here. Our results therefore suggest that analogous correlations between galaxy properties and local kinematic quantities could be explored in reconstructed velocity fields of the nearby Universe, extending cosmic web analyses beyond density-based classifications toward a more dynamical description of structure formation.

\section{Conclusions}
\label{sec:conclusions}

    We have investigated the connection between dark matter halo mass and the local cosmic velocity field within cosmic filaments in the TNG50-1-Dark simulation box at $z = 0$. Cosmic filaments were extracted using the 1-DREAM (1-Dimensional Recovery, Extraction, and Analysis of Manifolds) interpretable machine-learning framework and classified into three overdensity regimes to disentangle velocity-field-driven effects from density-driven ones. For each filament halo, we characterized the local velocity field through its divergence and vorticity, studying their correlations with halo mass and radial distance from filament spines. Our main findings are as follows:

\begin{itemize}
    \item[$\bullet$] Both divergence and vorticity display characteristic radial profiles within filaments, with enhanced amplitudes at intermediate distances from the spine ($\sim0.5{-}1.2\,{\rm Mpc}$), particularly in intermediate- and high-overdensity filaments, reflecting more complex dynamical environments.
    
    \item[$\bullet$] The most massive halos within filaments ($M \geq 10^{12}\,M_\odot$) exhibit strong spatial and kinematic segregation, preferentially residing near filament spines and in regions of simultaneously low absolute divergence and low vorticity. Quantitatively, $99.25\%$ and $98.87\%$ of these halos have $|\nabla \cdot \mathbf{v}| < 100\,\mathrm{km}\,\mathrm{s}^{-1}\,\mathrm{Mpc}^{-1}$ and $|\tilde{\omega}| < 100\,\mathrm{km}\,\mathrm{s}^{-1}\,\mathrm{Mpc}^{-1}$, respectively. Lower-mass halos span a much broader range of radial distances and kinematic states, dominating the most dynamically active environments. $\sim90\%$ of the total halo mass is likewise concentrated toward this low-activity region of phase space, with a mild preference for weakly converging flows.

    \item[$\bullet$] The kinematic constraints on halo mass persist across all three filament overdensity regimes, demonstrating that the velocity field carries information about halo mass independently of local density and provides a complementary axis of description for the cosmic web.
    
    \item[$\bullet$] Halos in kinematically calm environments assembled the bulk of their mass at later times and sustain higher recent accretion rates than halos in dynamically active regions, establishing the local kinematic state as a possible regulator of halo growth across cosmic time.
\end{itemize}
    
    Our results suggest that incorporating the local kinematics of filamentary flows into models of halo formation offers a promising path toward a more complete, dynamically informed description of mass assembly in the cosmic web, one that extends the traditional density-based picture of large-scale structure. Future work should test the robustness of these trends across different velocity-field reconstruction techniques and simulation suites with larger volumes or baryonic prescriptions, as tidal interactions and feedback processes may modify the results identified here. Extending the analysis to multiple redshift snapshots would further clarify whether the dependence on the velocity field components evolves with cosmic time or reflects a universal feature of accretion of matter onto filaments. On the observational side, PV estimates offer a promising avenue for testing these predictions in the nearby Universe, where the reconstructed divergence and vorticity fields could be coupled to galaxy formation and therefore act as a proxy of halo assembly.

\section*{Data Availability}
   
   The analysis presented in this work utilizes three publicly available resources: the filament identification framework 1-DREAM \citep{canducci20221dream}, with source code accessible via its GitLab repository\footnote{Toolbox \url{https://git.lwp.rug.nl/cs.projects/1DREAM}}; the IllustrisTNG halo group catalog \citep{nelson2019first}; and the IllustrisTNG halo structure catalog \citep{anbajagane2022}. The latter two are publicly available at \url{https://www.tng-project.org/}.

\begin{acknowledgements}
DDA, DR and HH acknowledge funding from the European Research Council (ERC) under the European Union’s Horizon 2020 research and innovation program (Grant agreement No. 101053992).
\end{acknowledgements}

\bibliographystyle{aa}
\bibliography{references}

\begin{appendix}
\section{Additional details of the 1-DREAM pipeline}

   \begin{figure}
   \centering
        \begin{subfigure}[b]{.45\textwidth} 
        \centering 
        \includegraphics[width=\hsize]{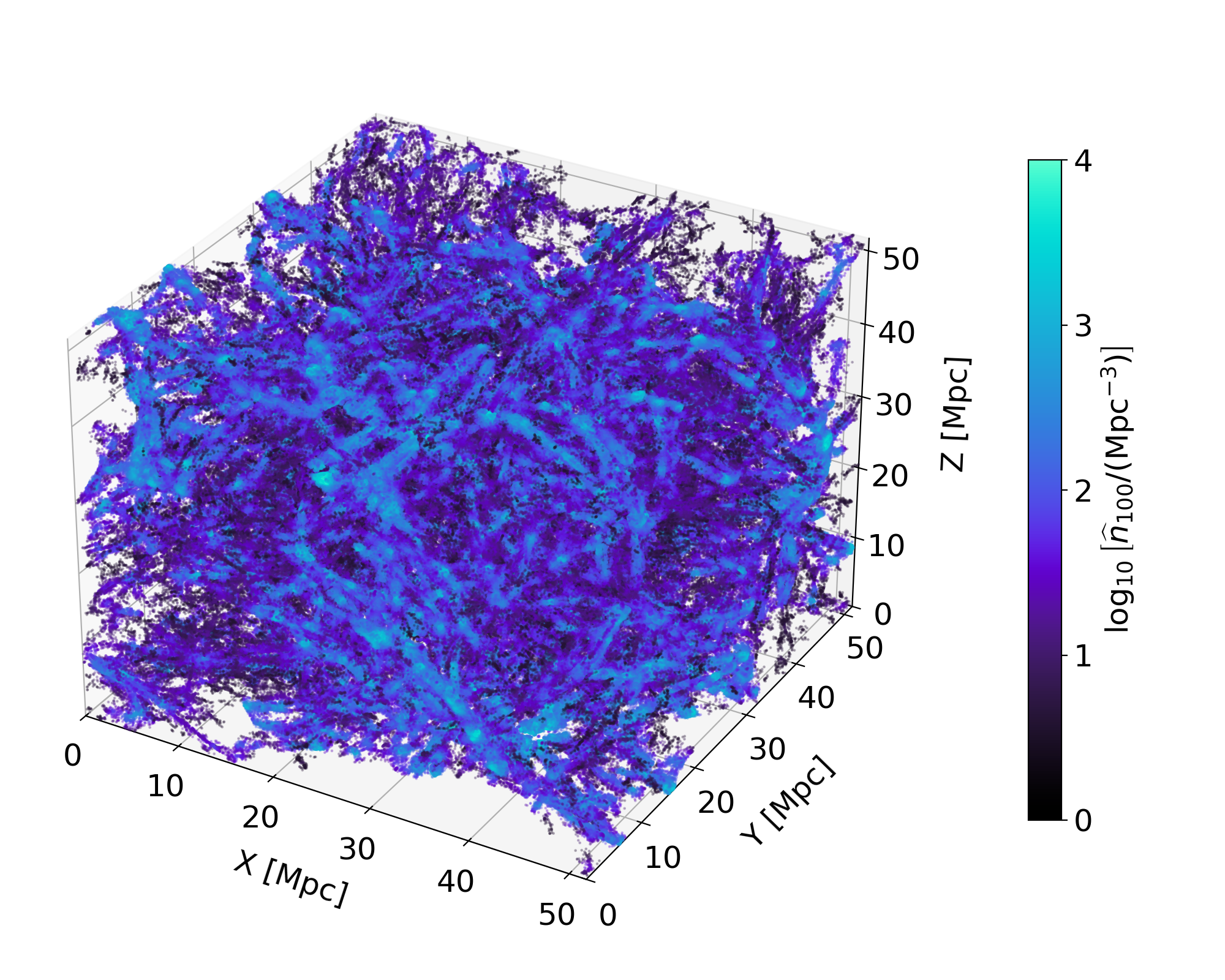} 
        \caption{Filaments}
        \label{fig:one_dimensional_partition}
        \end{subfigure}
        \begin{subfigure}[b]{.45\textwidth} 
        \centering 
        \includegraphics[width=\hsize]{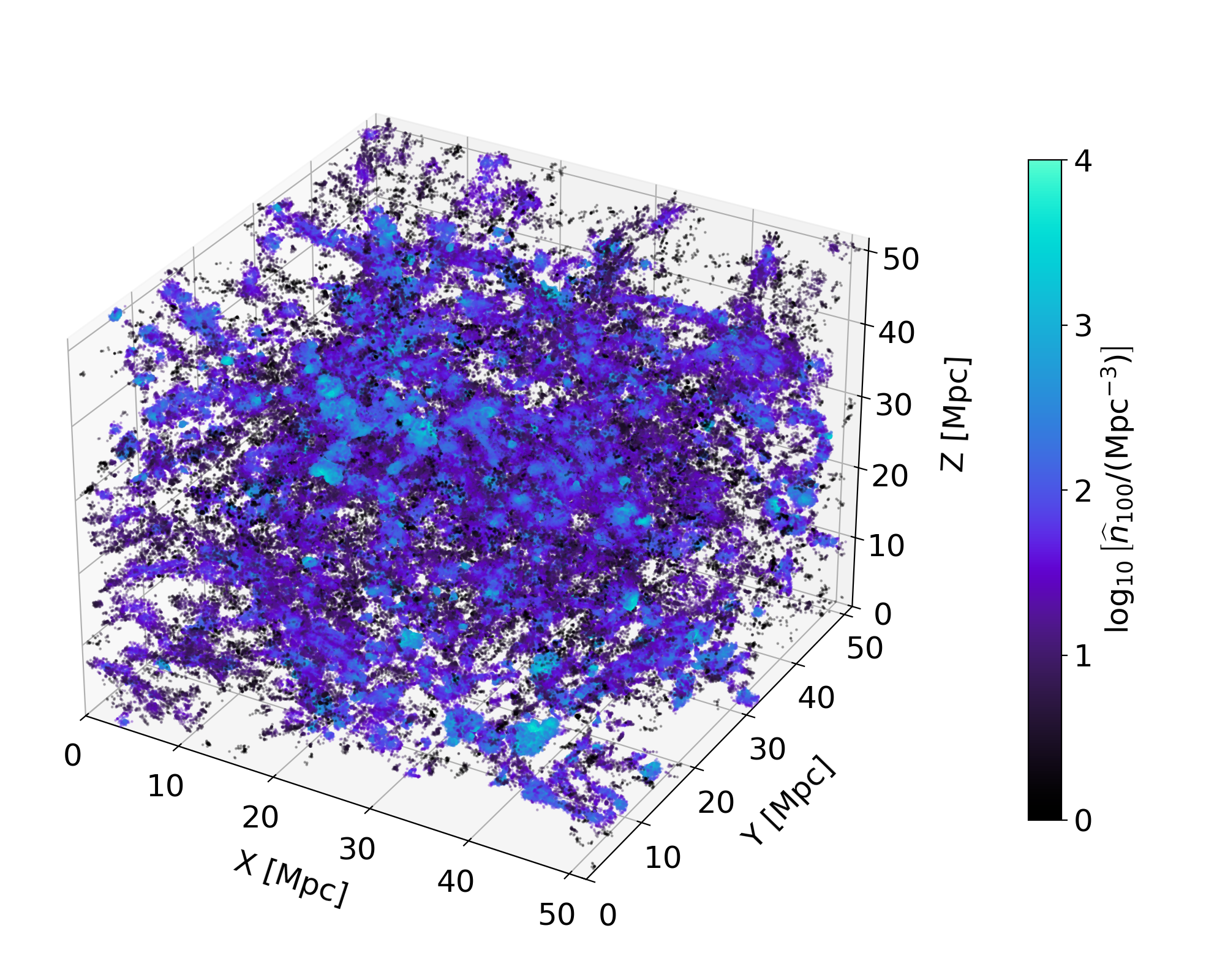} 
        \caption{Sheets}
        \end{subfigure}
        \begin{subfigure}[b]{.45\textwidth} 
        \centering 
        \includegraphics[width=\hsize]{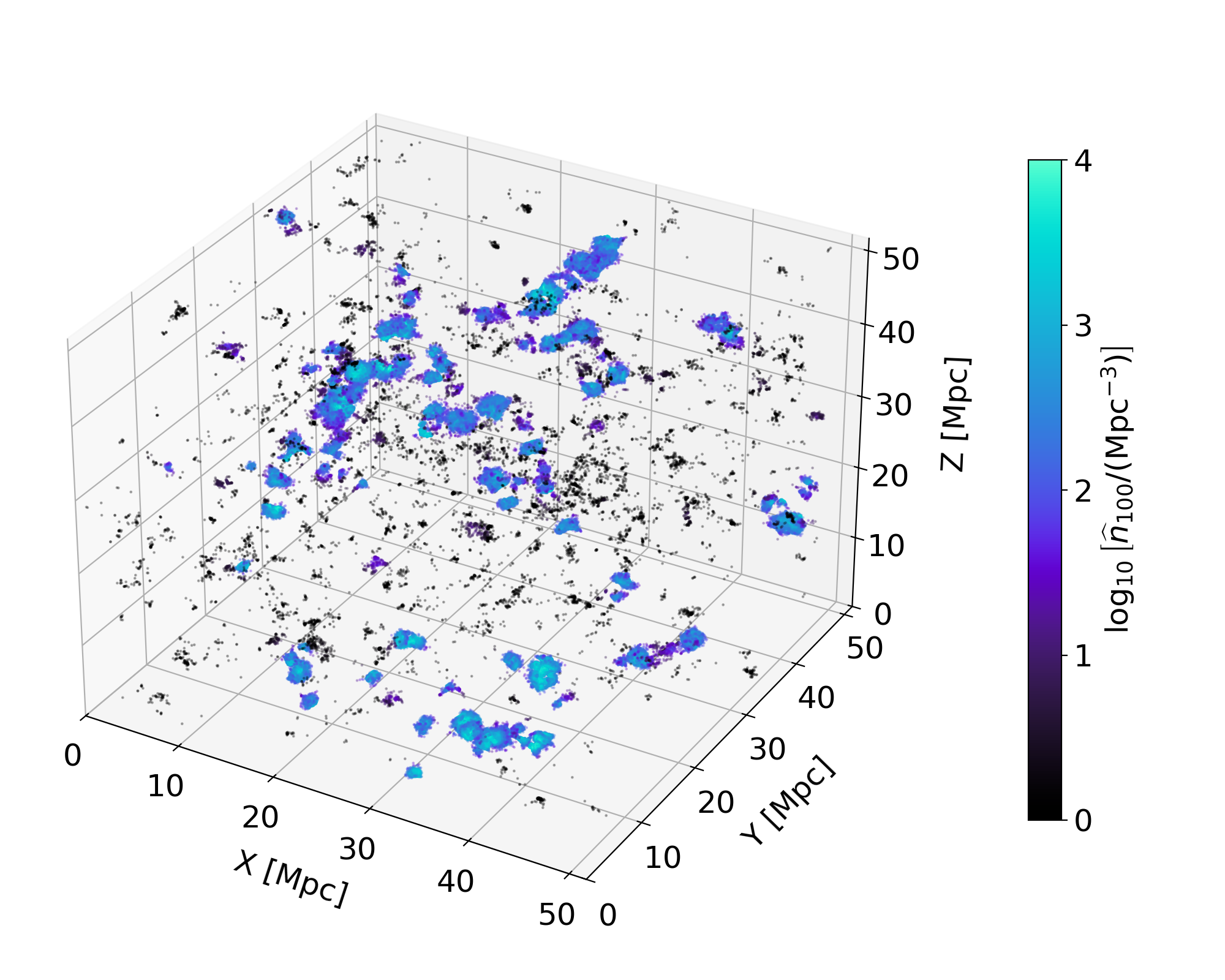} 
        \caption{Clusters}
        \end{subfigure}
      \caption{Three-dimensional distributions of halos classified into cosmic web environments as (a) filaments, (b) sheets, and (c) clusters, following the successive application of the 1-DREAM algorithms LAAT, MBMS, and DimIndex. Each point represents a DM halo, with color indicating the logarithm of its local $k$-nearest-neighbor halo number density, estimated using the distance to its 100th nearest neighbor within the corresponding structural class. Density values are clipped at the 1st and 99th percentiles for visualization.}
         \label{fig:laat_by_dimidx}
   \end{figure}

   Figure~\ref{fig:laat_by_dimidx} provides a visual overview of the cosmic-web classification obtained following the application of 1-DREAM's LAAT, MBMS, and DimIndex, showing the three-dimensional distribution of halos identified as belonging to filaments, sheets, and clusters. For visualization, we estimate the local halo number density using a $k$-nearest-neighbor estimator applied separately to each DimIndex-classified population. For every halo, we determine the distance ($r_k$) to its $k=100$-th neighboring halo and define $\widehat n_k=\frac{k}{(4\pi/3)r_k^3}$, with the logarithm of this quantity used to color the points in Fig.~\ref{fig:laat_by_dimidx}. This density estimate is used only to illustrate the internal spatial distribution of the three structural classes and is distinct from the filament overdensity defined in Section~\ref{sec:classification_methods}.

Table~\ref{table:1dream_full_params} lists the full set of parameters used in the 1-DREAM pipeline for this work. Each algorithm is described briefly in Section~\ref{sec:1dream_methods}; here we provide the specific values adopted for the filament extraction from the TNG50-1-Dark simulation.

\begin{table}
\caption{Full list of 1-DREAM parameters}
\label{table:1dream_full_params}
$$ 
     \begin{array}{p{0.2\linewidth}l p{0.1\linewidth}p{0.4\linewidth}}
        \hline
        \noalign{\smallskip}
        Algorithm & \mathrm{Parameter} & Value & Description \\
        \noalign{\smallskip} \hline \noalign{\smallskip}
        LAAT & \mathrm{r \in \mathbb{R}} & 0.5 & \textbf{Neighbourhood radius} \\
        & \mathrm{\zeta \in \mathbb{R}} & 0.05 & Evaporation rate \\
        & \mathrm{\kappa \in \mathbb{R}} & 0.8 & Shape v. pheromone \\
        & \mathrm{\omega \in \mathbb{R}} & 10 & Inverse temperature \\
        & \mathrm{\gamma \in \mathbb{R}} & 0.05 & Deposited pheromone \\
        & \mathrm{F \in \mathbb{R}} & 10 & Pheromone threshold \\
        & \mathrm{N_{epochs} \in \mathbb{N}} & 100 & Number of epochs \\
        & \mathrm{N_{steps} \in \mathbb{N}} & 12000 & Step per epochs \\
        & \mathrm{N_{ants} \in \mathbb{N}} & 200 & Number of agents \\
        \noalign{\smallskip} \hline \noalign{\smallskip}
        MBMS & \mathrm{r \in \mathbb{R}} & 1 & \textbf{Neighbourhood radius} \\
        & \mathrm{N_g \in \mathbb{N}} & 3 & Number of generations \\
        \noalign{\smallskip} \hline \noalign{\smallskip}
        Dim Index & \mathrm{r \in \mathbb{R}} & 0.5 & \textbf{Neighbourhood radius} \\
        & \mathrm{\tau \in \mathbb{N}} & 5 & Filtering threshold \\
        \noalign{\smallskip} \hline \noalign{\smallskip}
        Crawling & \mathrm{r \in \mathbb{R}} & 0.5 & \textbf{Neighbourhood radius} \\
        & \mathrm{\beta \in \mathbb{R}} & 0.3 & Jump tolerance \\
        \noalign{\smallskip} \hline \noalign{\smallskip}
        fixed-mean GMM & \mathrm{r \in \mathbb{R}} & 0.5 & \textbf{Neighbourhood radius} \\

        \hline
     \end{array}
 $$ 
\end{table}

    \section{Further comments on the followed methodology and analysis}

    Our analysis is limited to the TNG50-1-Dark simulation volume and therefore remains subject to finite-volume and resolution effects. The supplementary quantities used to probe halo growth histories in Section~\ref{sec:discussion} are furthermore restricted to halos above the minimum mass threshold of the IllustrisTNG Halo Structure catalog ($M \gtrsim 10^{8.8}\,M_\odot$ for TNG50), introducing incompleteness at the low-mass end. 
    Additionally, the thresholds used to define the low- and high-dynamical-activity subsamples in Figure~\ref{fig:mass_assembly} are not physically unique, but were selected to separate two different kinematic regimes. Note that the high-activity sample is not simply the complement of the low-activity sample. We exclude mixed cases such as $|\nabla\cdot\mathbf{v}|=150$, $|\boldsymbol{\omega}|=30\,\mathrm{km}\,\mathrm{s}^{-1}\,\mathrm{Mpc}^{-1}$ and $|\nabla\cdot\mathbf{v}|=50$, $|\boldsymbol{\omega}|=200\,\mathrm{km}\,\mathrm{s}^{-1}\,\mathrm{Mpc}^{-1}$. We repeated the analysis using dynamical-activity thresholds ranging from 50 to 200 $\mathrm{km}\,\mathrm{s}^{-1}\,\mathrm{Mpc}^{-1}$. In every case, halos in the low-activity subsample retained a lower median formation redshift and a higher median recent mass accretion rate than halos in the high-activity subsample. Although the precise median values and distribution widths varied slightly with each adopted threshold, the trends of both relations were preserved. Our conclusions are therefore not driven by the adopted dynamical-activity cuts. Future work should test the robustness of these trends across different cosmological simulations and larger volumes. Incorporating full halo evolutionary histories and baryonic physics would further clarify the physical origin of the observed kinematic correlations and their potential observational signatures.
    
    \subsection{Filament spine length}

   \begin{figure}
    \centering
    \includegraphics[width=\hsize]{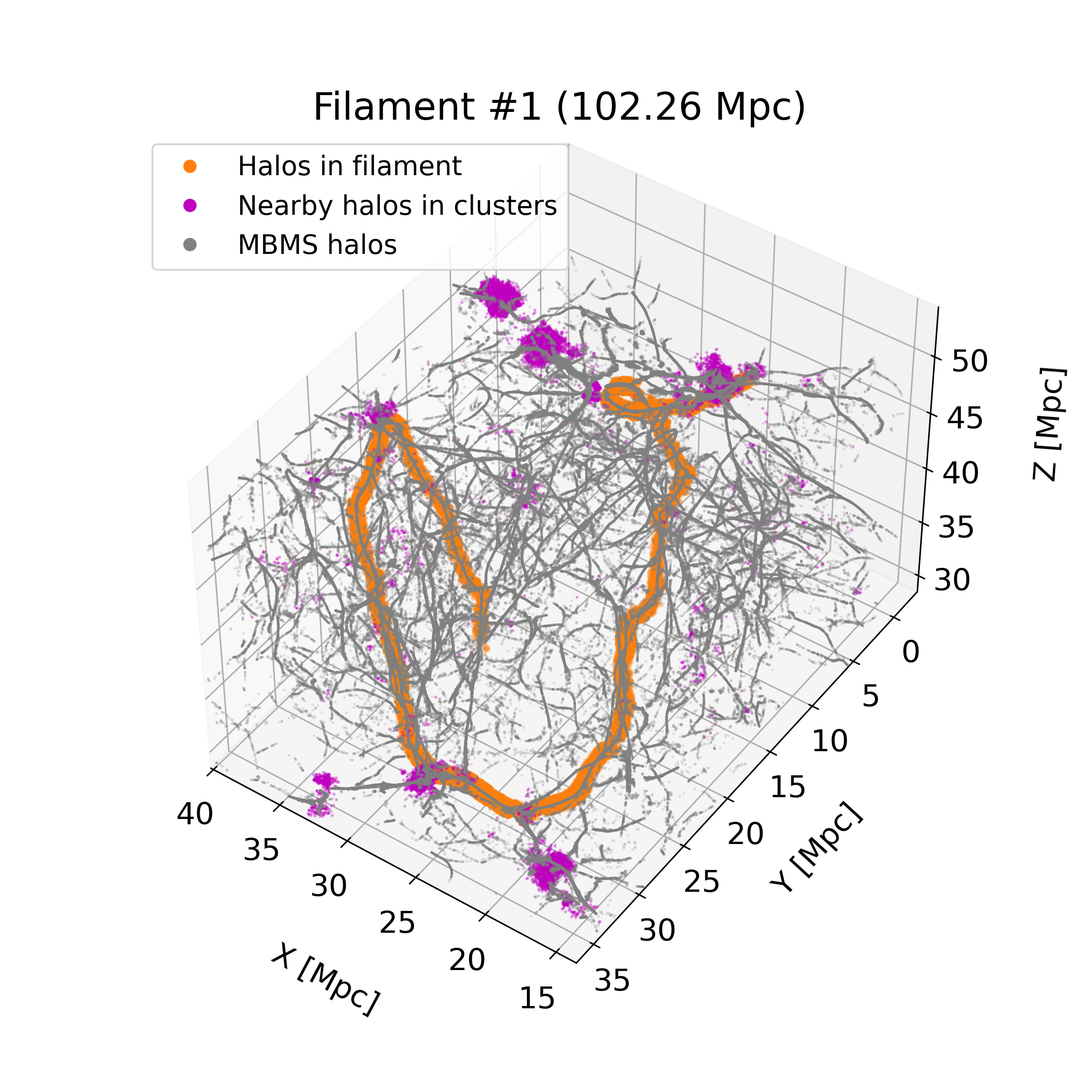}
    \includegraphics[width=\hsize]{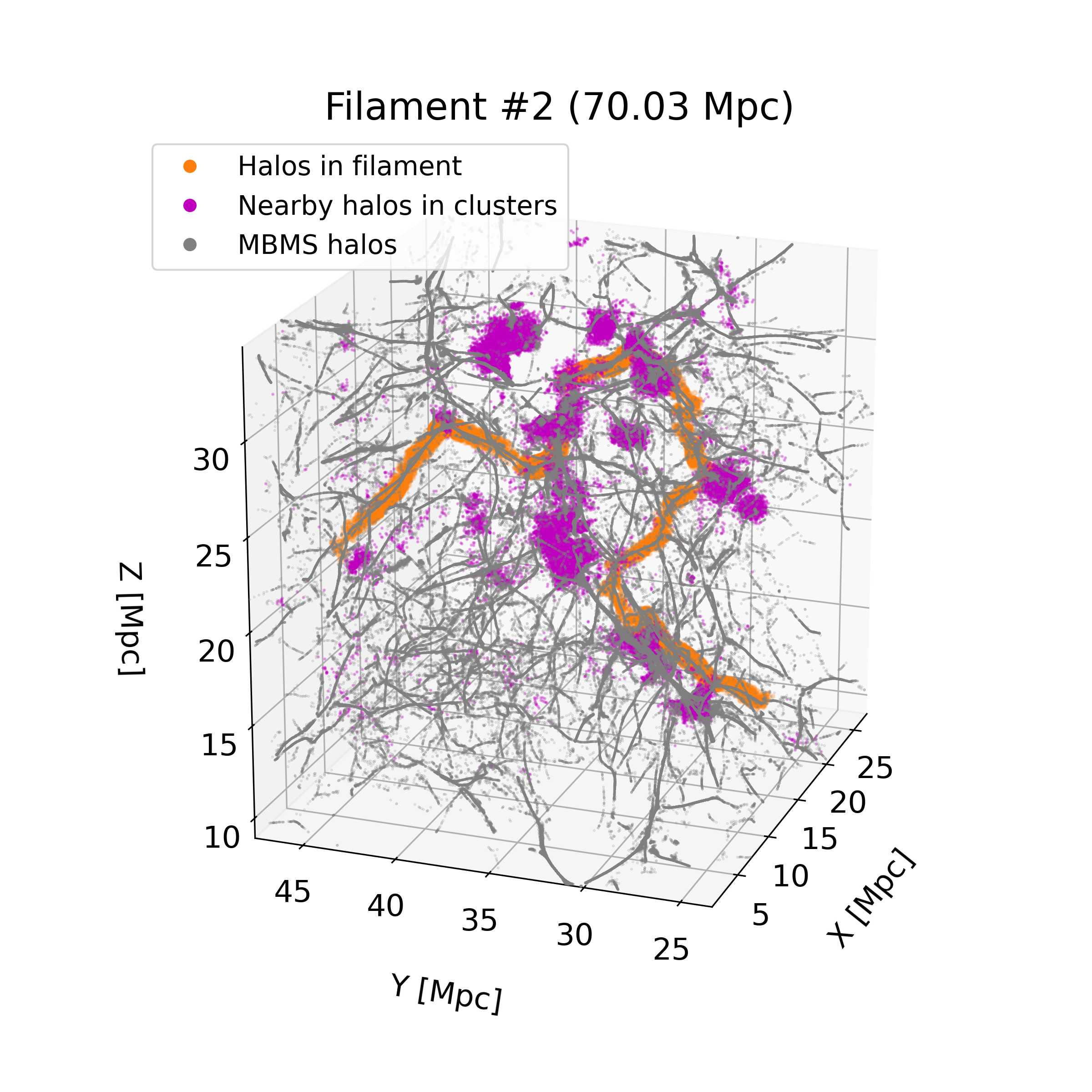}
      \caption{Three-dimensional view of the spatial distribution of halos around the two longest filaments in the catalog, filament 1 (top) and filament 2 (bottom). Orange points show halos belonging to each filament, magenta points show nearby halos belonging to cluster-like environments, and gray points show MBMS halos tracing the refined structures within each local subset.}
         \label{fig:longest_filaments}
   \end{figure}
    
    Although the simulation volume has a side length of $\sim 50$ Mpc, two retrieved filaments have total spine lengths larger than this value, as shown in the bottom panel of Fig.~\ref{fig:catalog_distribution}. This is possible because the reported length corresponds to the arclength of the reconstructed spine, rather than to the end-to-end spatial extent of the structure. As illustrated in Fig.~\ref{fig:longest_filaments}, filaments 1 and 2 have cumulative spine lengths of 102.26 and 70.03 Mpc, respectively, tracing curved, extended paths through the simulation box. Their large lengths are the result of MMCrawling maintaining continuity in the construction of the filament spine graphs through cluster-like environments, where the one-dimensional filamentary signal is partially interrupted or sparsely sampled. Ideally, MMCrawling should stop at this drop in local density and end the tracing of the corresponding filament. This is influenced by the jump tolerance parameter $\beta$ of MMCrawling, which can be calibrated to regulate the degree to which the algorithm connects separated one-dimensional halos into a continuous spine. This parametrization was however chosen to achieve the best characterization of the vast majority of the filaments in our sample. Given that our analysis is based on the halo population of filament environments independent of their lengths, inaccuracies in the measurements of filament lengths have no effect on our results.  

    \subsection{GMM training aperture and search radius}
    \label{app:gmm_training}
    As described in Section~\ref{sec:1dream_methods}, the probabilistic model of each filament is trained using halos within 0.5\,Mpc of the spine and subsequently evaluated to assign filament-membership likelihoods to halos out to 3\,Mpc. Since the trained aperture is substantially smaller than the radius over which the model is applied, we tested the sensitivity of our results to both choices.

\begin{figure*}
    \centering
        \begin{subfigure}[b]{.68\textwidth} 
        \centering 
        \includegraphics[width=\hsize]{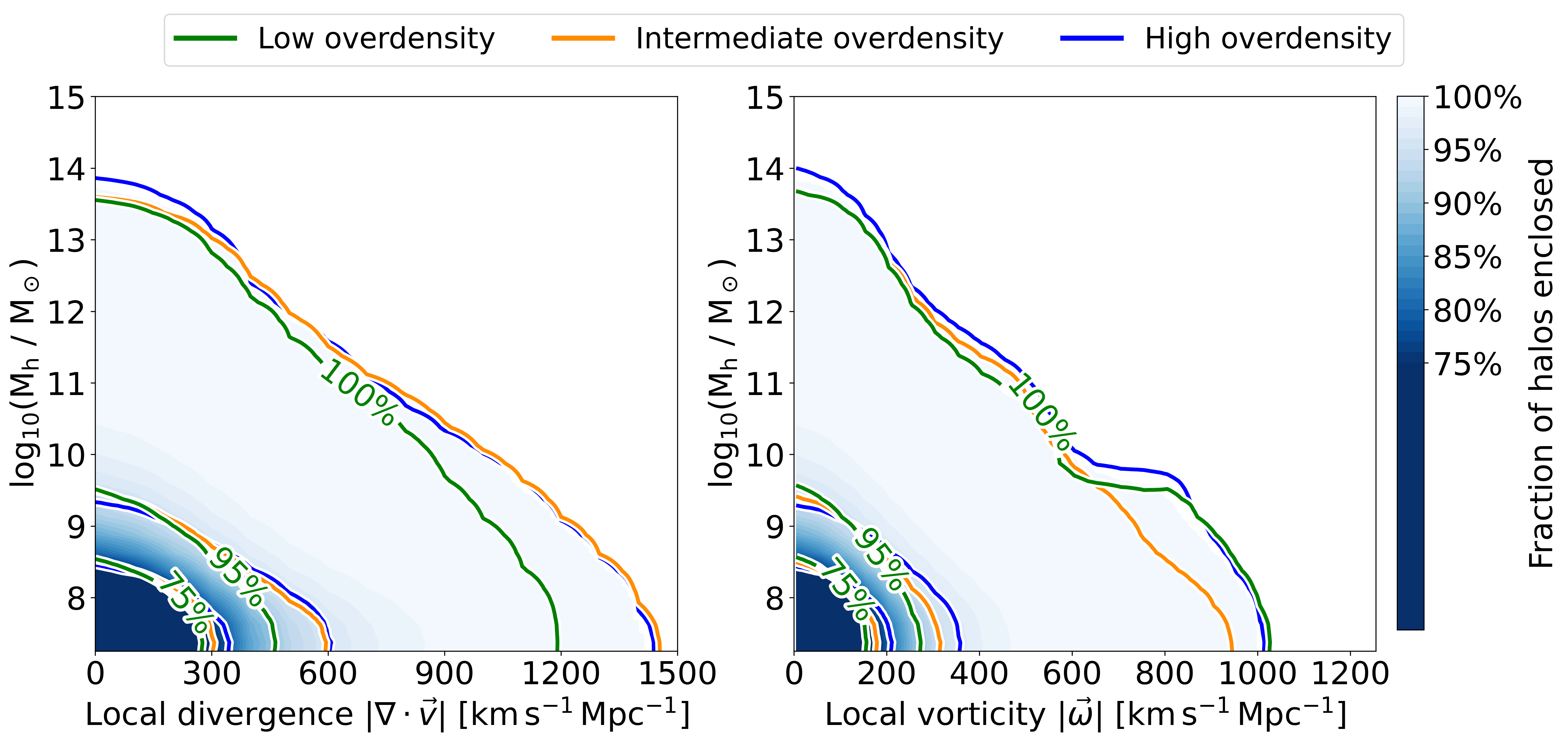} 
        \caption{$0.3\,\mathrm{Mpc}$ training aperture and $3\,\mathrm{Mpc}$ search radius}
        \label{fig:mass_new_gmm_a}
        \end{subfigure}
        \\
        \medskip
        \begin{subfigure}[b]{.68\textwidth} 
        \centering 
        \includegraphics[width=\hsize]{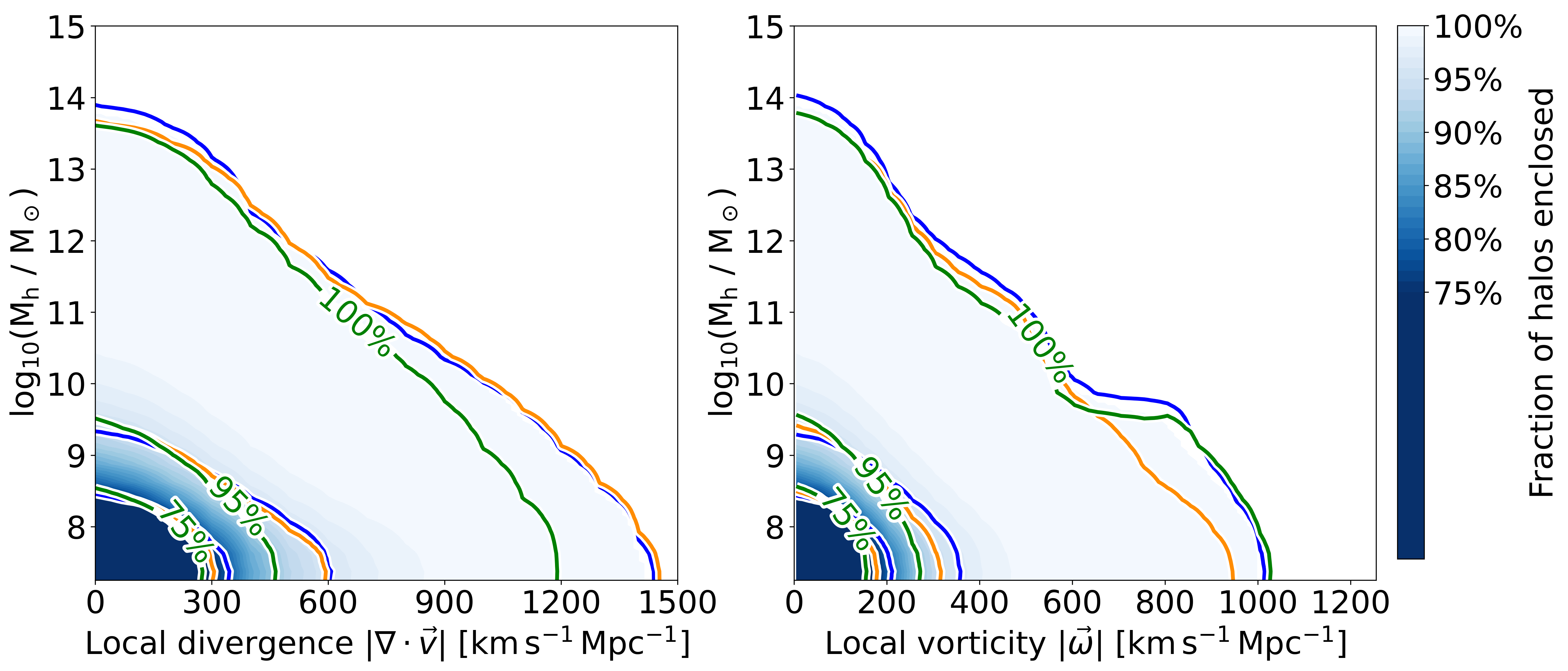} 
        \caption{$0.4\,\mathrm{Mpc}$ training aperture and $3\,\mathrm{Mpc}$ search radius}
        \label{fig:mass_new_gmm_b}
        \end{subfigure}
        \\
        \medskip
        \begin{subfigure}[b]{.68\textwidth} 
        \centering 
        \includegraphics[width=\hsize]{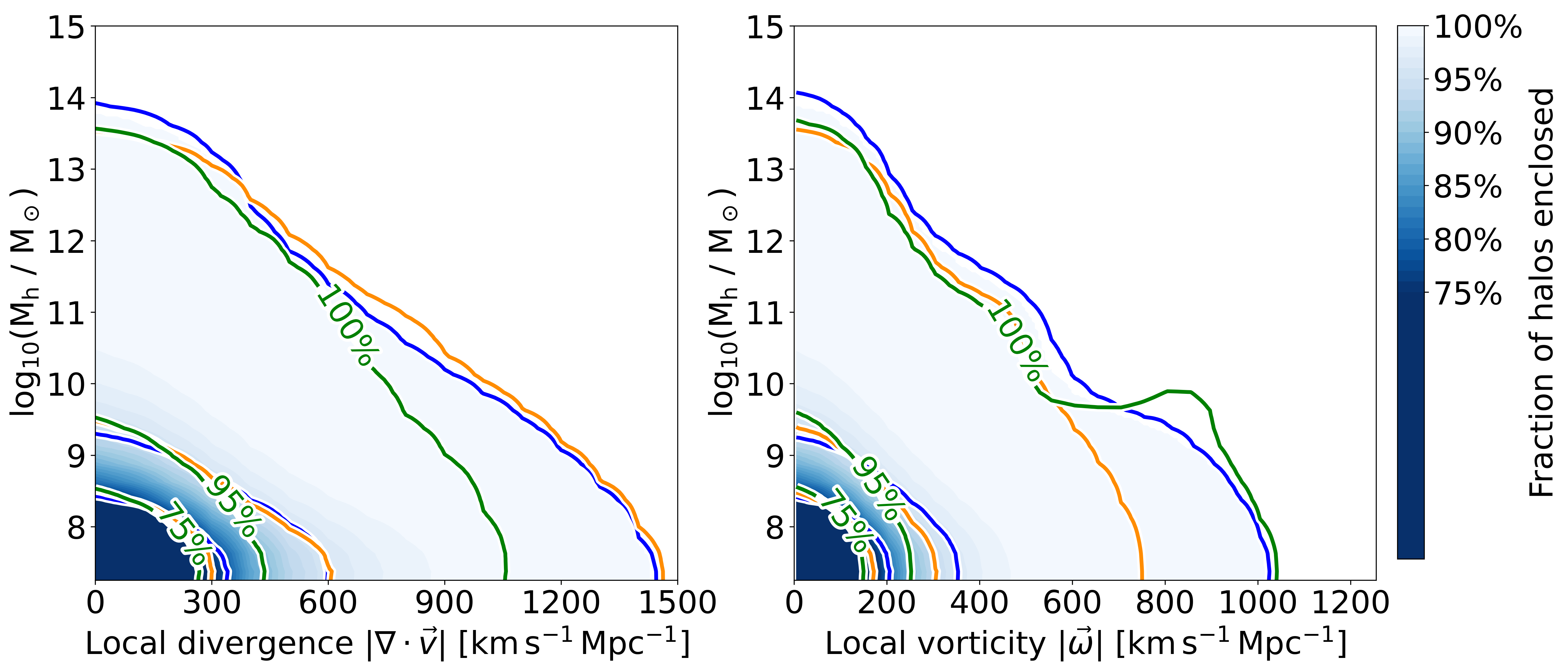} 
        \caption{$0.5\,\mathrm{Mpc}$ training aperture and $2\,\mathrm{Mpc}$ search radius}
        \label{fig:mass_new_gmm_c}
        \end{subfigure}
      \caption{Sensitivity of the halo mass-velocity field relation to the GMM training aperture and halo search radius. Panels show contour plots as seen in Figure~\ref{fig:mass_curl_div}, but reproduced for three combinations of training aperture and search radius: (a) 0.3\,Mpc and (b) 0.4\,Mpc training apertures, both with the fiducial 3\,Mpc search radius; and (c) the fiducial 0.5\,Mpc training aperture with a reduced 2\,Mpc search radius. Green, orange, and blue contours correspond to the low-, intermediate-, and high-overdensity filament samples, respectively.}
         \label{fig:mass_new_gmm}
   \end{figure*}

    We first varied the GMM training aperture while keeping the search radius fixed at 3\,Mpc, retraining the model using halos within 0.4 and 0.3\,Mpc of the spine. We could not test apertures larger than 0.5\,Mpc, as this was the maximum radius available from the MMCrawling output with the parameters adopted in this work. The resulting filament-halo catalogs are essentially unchanged. Relative to the fiducial sample of 2,792,885 halos, the total number of retained halos varies by less than $0.04\%$. The fraction of halos residing in converging flow also differs from the fiducial case by at most $0.11$ percentage points across all three overdensity categories. We additionally reproduced Figure~\ref{fig:mass_curl_div} for both alternative training apertures, shown in Figures~\ref{fig:mass_new_gmm_a} and \ref{fig:mass_new_gmm_b}. The enclosed-fraction contours remain essentially unchanged, indicating that the main halo mass-velocity field trends are robust to the adopted GMM training aperture within the range tested.

   \begin{figure*}
    \centering
        \begin{subfigure}[b]{.68\textwidth} 
        \centering 
        \includegraphics[width=\hsize]{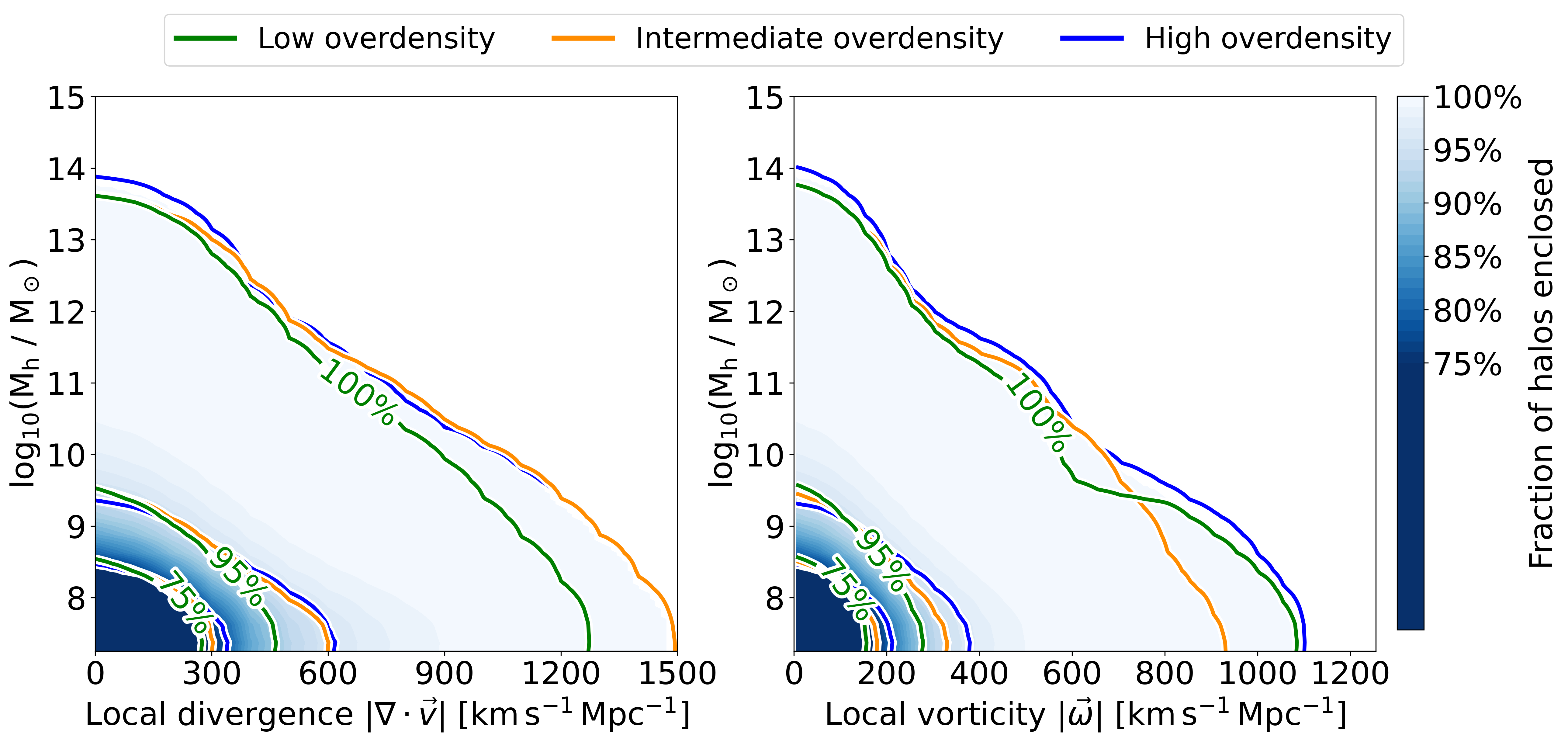} 
        \caption{Finer voxel resolution ($\Delta x = 0.2$ Mpc)}
        \end{subfigure}
        \\
        \medskip
        \begin{subfigure}[b]{.68\textwidth} 
        \centering 
        \includegraphics[width=\hsize]{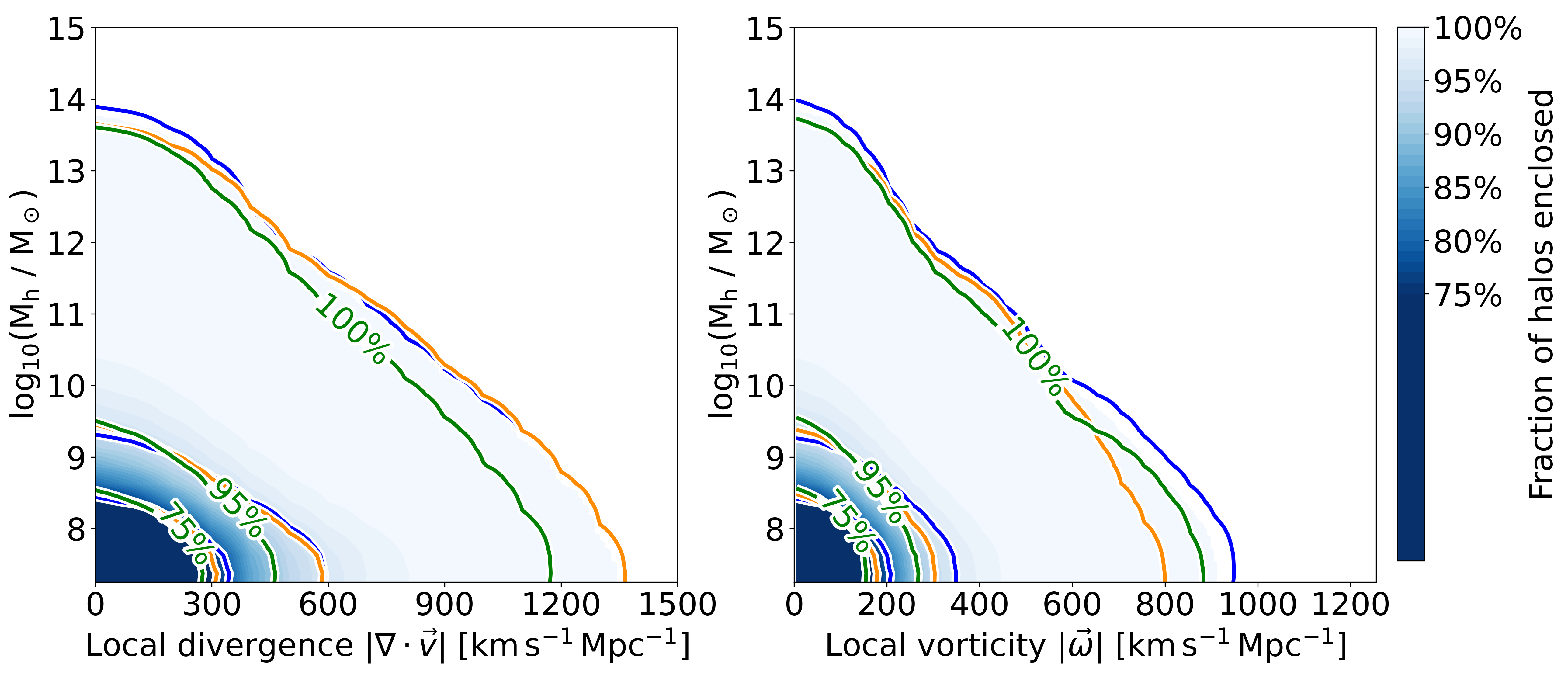} 
        \caption{Coarser voxel resolution ($\Delta x = 0.4$ Mpc)}
        \end{subfigure}
      \caption{Sensitivity of the halo mass-velocity field relation to the spatial resolution used to construct the velocity field. Panels show contour plots as seen in Figure~\ref{fig:mass_curl_div}, but reproduced for two different voxel resolutions: (a) $\Delta x = 0.2$ Mpc and (b) $\Delta x = 0.4$ Mpc. The Gaussian smoothing scale is kept fixed at $\sigma=0.4\,\mathrm{Mpc}$ in both cases. Green, orange, and blue contours correspond to the low-, intermediate-, and high-overdensity filament samples, respectively.}
         \label{fig:mass_new_res}
   \end{figure*}

    We separately tested the effect of the search radius by reducing it from 3 to 2\,Mpc while keeping the fiducial 0.5\,Mpc training aperture (Fig.~\ref{fig:mass_new_gmm_c}). This reduced the total number of retained halos from 2,792,885 to 1,855,538, corresponding to a decrease of $\sim34\%$. This is smaller than the $\sim56\%$ reduction expected from pure cylindrical volume scaling ($V \propto r^2$), reflecting that halos are preferentially concentrated toward the filament spine rather than uniformly distributed throughout the filament cross-section. Regarding the 75\% and 95\% contours, Figure~\ref{fig:mass_new_gmm_c} preserves the main mass–kinematic segregation seen in the fiducial sample. We notice that decreasing the search radius also changes the converging-flow fractions in a manner consistent with the radial trends reported in Section~\ref{subsec:radial_dist}. In low-, intermediate-, and high-overdensity filaments, the converging fraction changes from $65.24\%$, $80.97\%$, and $83.10\%$ to $64.15\%$, $83.34\%$, and $83.69\%$, respectively. This reflects the different radial behavior shown in Figure~\ref{fig:div_dist}, where low-overdensity filaments exhibit a gradual increase in converging-flow fraction with radial distance, while the intermediate- and high-overdensity samples decline again toward the outer $2$-$3\,\mathrm{Mpc}$ region. We therefore retain 3\,Mpc as our fiducial search radius, both to capture this outer radial behavior and because it is consistent with the typical radial extent of filaments discussed in Section~\ref{sec:1dream_methods}.

    \subsection{Velocity field grid resolution and smoothing}
    \label{app:resolution}
    The construction of the velocity field depends on the adopted grid resolution and Gaussian smoothing scale, which impact the magnitude and spatial structure of the derived divergence and vorticity fields. Our fiducial choice of $\Delta x = 0.3\,\mathrm{Mpc}$ is comparable to the mean inter-halo separation ($\sim 0.28\,\mathrm{Mpc}$) and was selected to balance spatial resolution and suppression of sampling noise. However, dense filament regions can contain many halos within a single voxel, potentially smoothing over small-scale variations in the derived velocity-gradient field. To test this, we repeated the full velocity-field construction, gradient calculations, and halo assignment using voxel resolutions of $\Delta x = 0.2$ and 0.4\,Mpc, while keeping the physical Gaussian smoothing scale fixed at $\sigma = 0.4\,\mathrm{Mpc}$ in all cases. These two alternative resolutions are shown in Figure ~\ref{fig:mass_new_res} and can be compared directly with the fiducial $\Delta x=0.3\,\mathrm{Mpc}$ results in Figure~\ref{fig:mass_curl_div}.

    We find a modest, monotonic dependence of the converging-flow fraction on voxel resolution, with finer voxels yielding slightly lower converging fractions. The effect is most pronounced in the high-overdensity category, where the converging fraction changes from $82.32\%$ at $\Delta x = 0.2\,\mathrm{Mpc}$ to $83.10\%$ at the fiducial resolution and $84.13\%$ at $\Delta x = 0.4\,\mathrm{Mpc}$. Across all three overdensity categories, the change relative to the fiducial resolution is at most $\sim1$ percentage point, substantially smaller than the differences in converging-flow fractions across overdensity categories reported in Section~\ref{sec:results}. The slight increase in converging fraction toward coarser resolution may reflect the averaging of the mass-weighted velocity field over a larger volume, which suppresses small-scale variations and opposing flows. The $75\%$ and $95\%$ enclosed-fraction contours in Figure~\ref{fig:mass_new_res} remain nearly identical in shape and extent to the fiducial results, indicating that the region of phase space occupied by the bulk of the halo population is relatively insensitive to voxel size over the range tested.

    The outer $100\%$ contour is more sensitive to resolution, as it is determined by the small number of halos occupying the most extreme divergence and vorticity values. It extends to larger absolute divergence and vorticity values at finer resolution and contracts at coarser resolution, while the relative ordering of the overdensity categories along this contour is not always preserved. This sensitivity of the distribution tails does not affect the region occupied by the bulk of the halo population, from which our main conclusions are drawn. We therefore conclude that the fiducial $\Delta x = 0.3\,\mathrm{Mpc}$ resolution provides a reasonable intermediate choice within the range tested, and that our main results are robust to the adopted voxel size.
    
\end{appendix}

\end{document}